\documentclass[prd,
superscriptaddress,
showkeys]{revtex4}

\usepackage[english]{babel}
\usepackage{amsmath,bm}
\usepackage{amssymb}
\usepackage{enumitem}
\usepackage{textcomp}
\usepackage{graphicx}
\usepackage{dsfont}
\usepackage{color}
\usepackage{hyperref}
\usepackage{mathrsfs}
\graphicspath{{Figures/}}

\definecolor{purple}{rgb}{0.8,0,0.6}

\begin{document}

\title{Backreaction of electromagnetic fields and the Schwinger effect \\ in pseudoscalar inflation magnetogenesis}

\author{O.O.~Sobol}
\affiliation{Institute of Physics, Laboratory for Particle Physics and Cosmology (LPPC), \'{E}cole Polytechnique F\'{e}d\'{e}rale de Lausanne (EPFL), CH-1015 Lausanne, Switzerland}
\affiliation{Physics Faculty, Taras Shevchenko National University of Kyiv, 64/13, Volodymyrska Str., 01601 Kyiv, Ukraine}
\author{E.V.~Gorbar}
\affiliation{Physics Faculty, Taras Shevchenko National University of Kyiv, 64/13, Volodymyrska Str., 01601 Kyiv, Ukraine}
\affiliation{Bogolyubov Institute for Theoretical Physics, 14-b, Metrologichna Str., 03680 Kyiv, Ukraine}

\author{S.I.~Vilchinskii}
\affiliation{Institute of Physics, Laboratory for Particle Physics and Cosmology (LPPC), \'{E}cole Polytechnique F\'{e}d\'{e}rale de Lausanne (EPFL), CH-1015 Lausanne, Switzerland}
\affiliation{Physics Faculty, Taras Shevchenko National University of Kyiv, 64/13, Volodymyrska Str., 01601 Kyiv, Ukraine}
\affiliation{D\'{e}partement de Physique Th\'{e}orique, Center for Astroparticle Physics, Universit\'{e} de Gen\`{e}ve, 1211 Gen\`{e}ve 4, Switzerland}

\date{\today}
%\pacs{04.62.+v, 98.80.Cq, 98.62.En}

\begin{abstract}
We study magnetogenesis in axionlike inflation driven by a pseudoscalar field $\phi$ coupled axially to the electromagnetic (EM) field $(\beta/M_{p})\phi F_{\mu\nu}\tilde{F}^{\mu\nu}$ with dimensionless coupling constant $\beta$. A set of equations for the inflaton field, scale factor, and expectation values of quadratic functions of the EM field is derived. These equations take into account the Schwinger effect and the backreaction of generated EM fields on the Universe expansion. It is found that the backreaction becomes important when the EM energy density reaches the value $\rho_{\rm EM}\sim (\sqrt{2\epsilon}/\beta)\rho_{\rm inf}$ ($\epsilon$ is the slow-roll parameter and $\rho_{\rm inf}$ is the energy density of the inflaton) slowing down the inflaton rolling and terminating magnetogenesis. The Schwinger effect becomes relevant when the electric energy density exceeds the value $\rho_{E}\sim \alpha_{\rm EM}^{-3} (\rho_{\rm tot}^{2}/M_{p}^{4})$, where $\rho_{\rm tot}=3H^{2}M_{p}^{2}$ is the total energy density and $\alpha_{\rm EM}$ is the EM coupling constant. For large $\beta$, produced charged particles could constitute a significant part of the Universe energy density even before the preheating stage.
Numerically studying magnetogenesis in the $\alpha$-attractor model of inflation, we find that it is possible to generate helical magnetic fields with the maximal strength $10^{-15}\,{\rm G}$, however, only with the correlation length of order $1\,{\rm pc}$ at present.
\end{abstract}


\maketitle

\section{Introduction}
\label{sec-intro}

Recently magnetic fields with very large coherence scale $\lambda_{B}\gtrsim 1\,$Mpc were detected in cosmic voids through the gamma-ray observations of distant blazars \cite{Neronov:2010,Tavecchio:2010,Taylor:2011,Dermer:2011,Caprini:2015}. Combining the corresponding results with the data of the cosmic microwave background (CMB) \cite{Planck:2015-pmf,Sutton:2017,Jedamzik:2018,Paoletti:2018,Giovannini:2018b} and ultra-high-energy cosmic rays \cite{Bray:2018} observations constrains the strength of these fields to $10^{-18}\lesssim B\lesssim 10^{-9}\,$G.
It is well known also that magnetic fields are present on much smaller scales in the Universe \cite{Kronberg:1994,Grasso:2001,Widrow:2002,Giovannini:2004,Kandus:2011,Durrer:2013,Subramanian:2016,Giovannini:2018b} including galaxies and clusters of galaxies.
Observed intergalactic magnetic fields may be of either astrophysical or primordial origin and both magnetogenesis scenarios are currently under active consideration and study.
Although astrophysical mechanisms based on the Biermann battery \cite{Biermann:1950,Pudritz:1989,Gnedin:2000} have been proposed to generate the ``seed'' magnetic fields and various types of dynamo can enhance them \cite{Zeldovich:1980book,Lesch:1995,Kulsrud:1997,Colgate:2001}, it is problematic to use these mechanisms for the generation of magnetic fields with very large correlation length in cosmic voids indicating on the primordial origin of these magnetic fields. Clearly, this opens an intriguing possibility to obtain important information on the physical processes in the early Universe and poses the problem of how these fields were generated.

One of the natural mechanisms of primordial magnetic field generation is connected to the phase transitions in the early Universe. However, although
they may lead to magnetic fields of the necessary strength \cite{Hogan:1983,Quashnock:1989,Vachaspati:1991,Cheng:1994,Sigl:1997,Ahonen:1998}, the coherence length is determined by the horizon size at the moment of phase transition and is much less than Mpc today. On the other hand, the inflationary magnetogenesis \cite{Turner:1988,Ratra:1992} can provide the necessary seeds for the observed magnetic fields and attain very large coherence length.

In order to generate electromagnetic (EM) fields during inflation, the conformal invariance of Maxwell's action should be broken because fluctuations of the EM field are not enhanced in conformally flat inflationary background \cite{Parker:1968}. The conformal invariance can be broken by introducing the interaction of EM fields with scalar or pseudoscalar inflaton field or with the curvature scalar (see the pioneering works of Refs.~\cite{Turner:1988,Ratra:1992,Garretson:1992,Dolgov:1993}).  

Inflationary magnetogenesis typically faces a backreaction problem which can put strong constraints on the parameter space of the model \cite{Demozzi:2009}. Usually in the study of inflationary magnetogenesis, one solves first the equations governing the evolution of the inflaton field and the scale factor and then determines generated EM fields. However, such an approach is possible only if generated fields are weak enough not to influence the background evolution. In the opposite case, one has to take into account generated EM fields self-consistently and solve all equations of motion simultaneously. Clearly, this makes the problem of inflationary magnetogenesis much more complicated. Moreover, the backreaction of EM fields can modify the inflaton dynamics, change the Universe expansion, and make an impact on the spectrum of primordial perturbations. In our previous work \cite{Sobol:2018}, we considered the case of kinetic coupling between the scalar inflaton field and the EM field introduced by Ratra \cite{Ratra:1992} and revisited many times in the literature \cite{Giovannini:2001,Bamba:2004,Martin:2008,Demozzi:2009,Kanno:2009,Ferreira:2013,Ferreira:2014,Vilchinskii:2017,Sharma:2017b,Shtanov:2018} and showed that the backreaction of generated electric fields is indeed very important because it strongly suppresses magnetogenesis.

Because of the Schwinger effect \cite{Schwinger:1951} the generated electric fields can spontaneously create from vacuum the pairs of charged particles. The simplest case of pair creation in a constant and homogeneous background electric field in de Sitter space-time was investigated by many authors 
~\cite{Kobayashi:2014,  Froeb:2014,  Bavarsad:2016, Stahl:2016a, Stahl:2016b, Hayashinaka:2016a,  Hayashinaka:2016b,  Sharma:2017, Tangarife:2017, Bavarsad:2018, Hayashinaka:2018, Hayashinaka:thesis, Stahl:2018, Geng:2018, Bavarsad:2018}, where different space-time dimensions and the various types of charge carriers were considered.  
However, the case of a constant electric energy density is unphysical because maintaining this regime would require the existence of \textit{ad hoc} currents that could violate the second law of thermodynamics \cite{Giovannini:2018a}. Nevertheless, the expressions for the Schwinger current derived in this approximation have the same functional dependence as in a time-dependent electric background in the strong-field regime \cite{Kitamoto:2018}.

It should be noted that the cosmological Schwinger effect contains interesting features \cite{Kobayashi:2014,Hayashinaka:2016a,Hayashinaka:2016b,Hayashinaka:2018,Hayashinaka:thesis} that are absent in its flat-space counterpart such as (i) the infrared hyperconductivity in the bosonic case when the conductivity becomes very large in the limit of a small mass of charged particles and (ii) the negative conductivity in the weak field regime $eE\ll H^{2}$  which can, in principle, lead to the enhancement of the electric field \cite{Stahl:2018}. However, the definition of a finite part of the induced current is ambiguous and depends on the subtraction scheme. Thus, the negative conductivity is rather speculative \cite{Hayashinaka:2018,Hayashinaka:thesis} and may be an artifact of the used subtraction scheme. Moreover, in Ref.~\cite{Banyeres:2018} it was shown that the negative contribution to the conductivity is due to the nonlinear corrections to the Maxwell action and the logarithmic running of the coupling constant. Therefore, it cannot lead to any enhancement. 
	 
In this paper we consider the axial coupling of the EM field to a pseudoscalar inflaton field via the term $I(\phi)F_{\mu\nu}\tilde{F}^{\mu\nu}$ (here $\tilde{F}^{\mu\nu}$ is the dual EM tensor) taking into account the Schwinger effect and the backreaction of generated EM fields on the background evolution. In the case of the kinetic coupling model~\cite{Kanno:2009,Sobol:2018}, the electric component dominates the EM energy density causing strong backreaction and spoiling magnetogenesis. Therefore, the kinetic coupling model is unfavorable in contrast to pseudoscalar inflation \cite{Durrer:2011}. Magnetogenesis in the latter model is promising because it generates helical magnetic fields \cite{Garretson:1992,Durrer:2011,Anber:2006,Anber:2010,Barnaby:2012,Caprini:2014,Anber:2015,Ng:2015,Adshead:2015,Adshead:2016,Shtanov:2019} which are more stable against dissipation and, in addition, their coherence length can be further increased due to the inverse cascade process \cite{Joyce:1997,Cornwall:1997,Boyarsky:2012,Sydorenko:2016,Boyarsky:2015,Gorbar:2016a,Gorbar:2016b} in the chiral primordial plasma.
Since electric fields are also generated in this model, the pair production due to the Schwinger effect may become important and affect magnetogenesis. The same applies to the backreaction problem due to generated EM fields \cite{Garretson:1992,Durrer:2011,Anber:2006,Anber:2010,Barnaby:2012,Caprini:2014,Anber:2015,Ng:2015,Adshead:2015,Adshead:2016,Fujita:2015,Domcke:2018,Notari:2016,Figueroa:2018,Durrer:2019}. This means that both the Schwinger effect and
backreaction of EM fields should be taken into account in the study of magnetogenesis in pseudoscalar inflation.

Since the pseudoscalar inflation model looks promising (pseudoscalar field may appear also as a Kalb-Ramond axion in string theory inspired models \cite{Basilakos:2019a,Basilakos:2019b}), this provides the main motivation to extend our approach developed in 
\cite{Sobol:2018} to the case of pseudoscalar inflation magnetogenesis. Since it is
quite costly computationally to solve directly the equations for the EM field for each mode, our approach in Ref.~\cite{Sobol:2018} 
was connected with the derivation of a self-consistent set of equations for the electric field energy density, scale factor, and 
inflaton field taking into account the pair production due to the Schwinger effect. Perhaps 
not surprisingly, the implementation of this approach for magnetogenesis in the pseudoscalar inflation model turned out to be quite difficult. First of
all, since the axial coupling involves the scalar product $\mathbf{E}\cdot\mathbf{B}$ of electric and magnetic fields, it is immediately clear 
that the function $\mathbf{E}\cdot\mathbf{B}$ should
be necessarily taken into account in addition to the electric field energy density $\rho_E =\mathbf{E}^2/2$. Unfortunately, this strongly complicates
the analysis because then the self-consistency of equations for the EM field requires the introduction of new quadratic functions 
of electric and magnetic fields with additional spatial curls. Therefore, the corresponding system of equations governing the evolution of the EM field is not closed, and one has to use certain physically reasonable approximations which make it possible to close the system.

This paper is organized as follows. We formulate a self-consistent system of equations that govern the joint evolution of the scale factor, inflaton field, and quadratic functions of the EM field in Sec.~\ref{sec-basics}, taking into account the backreaction of generated fields and the Schwinger effect. In Sec.~\ref{sec-quant}, we derive the explicit form of the boundary terms in the Maxwell equations which describe the quantum-to-classical transition for modes undergoing the tachyonic instability during inflation. Section~\ref{sec-initial} is devoted to the determination of initial conditions for the EM field which should be imposed far from the end of inflation in order to study the system numerically. In Sec.~\ref{sec-Schwinger} we analyze the conditions when the Schwinger effect becomes important and affects magnetogenesis.  Section~\ref{sec-numerical} explains the approximation scheme which allows one to truncate the infinite system of equations and contains the numerical results in the $\alpha$-attractor inflationary model. The postinflationary evolution of generated magnetic fields is described in Sec.~\ref{sec-postinflationary}, where the present day values of the magnetic field strength and correlation length are determined. Finally, the summary of our work is given in Sec.~\ref{sec-concl}. Some technical details of the derivation of the initial conditions are given in Appendix~\ref{App-math}. We use in this paper the natural units and set $\hbar=c=1$.

\section{Evolution equations}
\label{sec-basics}

We consider a spatially flat Friedmann-Lema\^{i}tre-Robertson-Walker (FLRW) Universe with metric
\begin{equation}
\label{metric}
g_{\mu\nu}={\rm diag}\,(1,\,-a^{2},\,-a^{2},\,-a^{2}), \quad \sqrt{-g}=a^{3}.
\end{equation}
In order to study magnetogenesis in pseudoscalar inflation, we use the following action
for the inflaton field $\phi$ and the EM 4-potential $A_{\mu}$:
\begin{equation}
\label{action}
S=\int d^{4}x \sqrt{-g}\left[\frac{1}{2}\partial_{\mu}\phi\partial^{\mu}\phi-V(\phi)-\frac{1}{4}F_{\mu\nu}F^{\mu\nu}-\frac{1}{4}I(\phi)F_{\mu\nu}\tilde{F}^{\mu\nu}+\mathcal{L}_{\rm charged}(A_{\nu},\chi)\right],
\end{equation}
where $V(\phi)$ is the effective potential of the inflaton, $I(\phi)$ is a coupling function, and $\tilde{F}^{\mu\nu}=\frac{1}{2}\eta^{\mu\nu\lambda\rho}F_{\lambda\rho}$ is the dual EM tensor. In addition, $\eta^{\mu\nu\lambda\rho}=\varepsilon^{\mu\nu\lambda\rho}/\sqrt{-g}$, where $\varepsilon^{\mu\nu\lambda\rho}$ is the totally antisymmetric Levi-Civita symbol with $\varepsilon^{0123}=+1$ and $\mathcal{L}_{\rm charged}(A_{\nu},\chi)$ is a gauge-invariant Lagrangian of a charged field $\chi$ interacting with the EM field $A_{\mu}$.
The charged field $\chi$ is a generic field which could stand for a boson or fermion field as well as for any collection of them. It can, in principle, interact also with the inflaton directly. This interaction is crucial during the preheating stage because it determines the production of charged particles by the quickly oscillating inflaton. However, particle production is not effective during the slow-roll inflation because the inflaton field changes adiabatically slowly. Since the preheating stage lies beyond the scope of our paper, we will neglect the interaction between $\chi$ and $\phi$. 

Since $F_{\mu\nu}\tilde{F}^{\mu\nu}$ is a pseudoscalar, in order to preserve the parity symmetry the coupling function of the pseudoscalar inflaton field $\phi$ with $F_{\mu\nu}\tilde{F}^{\mu\nu}$ should be odd, $I(-\phi)=-I(\phi)$. In this paper, we consider 
the simplest axial coupling function linear in $\phi$
\begin{equation}
\label{coupling-function}
I(\phi)=\beta\frac{\phi}{M_{p}},
\end{equation} 
where $M_{p}=(8\pi G)^{-1/2}\approx 2.4\times 10^{18}\,{\rm GeV}$ is the reduced Planck mass and $\beta$ is a dimensionless coupling constant.

The Euler-Lagrange equations corresponding to action (\ref{action}) and the Bianchi identity have the form
\begin{equation}
\label{KGF}
\frac{1}{\sqrt{-g}}\partial_{\mu}\left[\sqrt{-g}g^{\mu\nu}\partial_{\nu}\phi\right]+\frac{dV}{d\phi}=-\frac{1}{4}I'(\phi)F_{\mu\nu}\tilde{F}^{\mu\nu},
\end{equation}

\begin{equation}
\label{Maxwell}
\frac{1}{\sqrt{-g}}\partial_{\mu}\left[\sqrt{-g}F^{\mu\nu}\right]+I'(\phi)\tilde{F}^{\mu\nu}\partial_{\mu}\phi=-j^{\nu},
\end{equation}

\begin{equation}
\label{Maxwell-Bianchi}
\frac{1}{\sqrt{-g}}\partial_{\mu}\left[\sqrt{-g}\tilde{F}^{\mu\nu}\right]=0,
\end{equation}
where
\begin{equation}
\label{4-current}
j^{\mu}=\frac{\partial\mathcal{L}_{\rm charged}(A_{\nu},\chi)}{\partial A_{\mu}}=(n,\, \mathbf{j})
\end{equation}
is the electric four-current. The right-hand side of Eq.~(\ref{KGF}) could be neglected if the generated fields are weak. 
If they are not, then the backreaction of these fields on the background evolution should be taken into account and a self-consistent system of equations 
should be solved.

According to the first Friedmann equation, the expansion rate of the FLRW Universe is described by the 00 component of the stress-energy tensor.
In the case under consideration, the latter is given by
\begin{equation}
T_{\mu\nu}=\frac{2}{\sqrt{-g}}\frac{\delta S}{\delta g^{\mu\nu}}=\partial_{\mu}\phi\partial_{\nu}\phi-g^{\lambda\rho}F_{\mu\lambda}F_{\nu\rho}
-g_{\mu\nu} \mathcal{L}_{0}+T_{\mu\nu}^{({\rm charged})},
\end{equation}
where $\mathcal{L}_{0}=\tfrac{1}{2} (\partial\phi)^{2}-V(\phi)-\tfrac{1}{4} F_{\mu\nu}F^{\mu\nu}$ and $T_{\mu\nu}^{({\rm charged})}$ describes the contribution due to 
the charged field $\chi$. Notice also that the axial coupling term does not contribute to the energy-momentum tensor because it does not depend on metric.

It is convenient to use the Coulomb gauge for the EM field $A_{\mu}=(0,\,\mathbf{A})$ where ${\rm div\,}\mathbf{A}=0$. Then electric 
and magnetic fields are defined as follows:
\begin{equation}
\label{fields-E-and-B}
\mathbf{E}=-\frac{1}{a}\dot{\mathbf{A}}, \qquad \mathbf{B}=\frac{1}{a^{2}} {\rm rot\,}\mathbf{A},
\end{equation}
where $a(t)$ is a scale factor of the FLRW background. The components of the EM tensors are expressed through the 
electric and magnetic fields as usual
\begin{equation}
\label{EM-tensors}
F^{0i}=\frac{1}{a} E^{i},\quad F_{ij}=a^{2}\varepsilon_{ijk}B^{k},
\quad \tilde{F}^{0i}=\frac{1}{a}B^{i}, \quad \tilde{F}_{ij}=-a^{2}\varepsilon_{ijk}E^{k},
\end{equation}
where $\varepsilon_{ijk}$ is the three-dimensional Levi-Civita symbol and the Latin indices $i,\,j,\,k$ on the right-hand side denote the components of
3-vectors. We would like to note that the electric and magnetic fields defined in Eqs.~(\ref{fields-E-and-B}) and (\ref{EM-tensors}) are physical fields measured by a comoving observer.

We will assume that the inflaton field is spatially homogeneous (the question whether perturbations of metric and the inflaton 
can always be neglected in the study of magnetogenesis is important and deserves a separate investigation). Then the energy density reads as
\begin{equation}
\label{energy-density}
\rho=\left[\frac{1}{2}\dot{\phi}^{2}+V(\phi)\right]+\frac{1}{2}\langle E^{2}+B^{2}\rangle+\rho_{\chi}=\rho_{\rm inf}+\rho_{\rm EM}+\rho_{\chi},
\end{equation}
where $\rho_{\rm inf}$ and $\rho_{\chi}$ are the energy densities of the inflaton and charged particles produced by the Schwinger effect, respectively. Since EM fields are generated through the enhancement of quantum fluctuations, their energy density $\rho_{\rm EM}=\frac{1}{2}\langle E^{2}+B^{2}\rangle$ is given by the expectation value of the corresponding operator.

The Friedmann, Klein-Gordon-Fock (KGF), and Maxwell equations define the following closed system of dynamic equations:
\begin{eqnarray}
\label{Friedmann}
&&H^{2}=\frac{1}{3M_{p}^{2}}\left(\rho_{\rm inf}+\rho_{\rm EM}+\rho_{\chi}\right), \\
\label{KGF-2}  
&&\ddot{\phi}+3H\dot{\phi}+\frac{dV}{d\phi}=I'(\phi)\left<\mathbf{E}\cdot\mathbf{B}\right>,\\
\label{Maxwell-2}
&&\dot{\mathbf{E}}+2H\mathbf{E}-\frac{1}{a}{\rm rot\,}\mathbf{B}+I'(\phi)\dot{\phi}\mathbf{B}=-a\mathbf{j},
\\
\label{Maxwell-Bianchi-2}
&&\dot{\mathbf{B}}+2H\mathbf{B}+\frac{1}{a}{\rm rot\,}\mathbf{E}=0,\\
&&{\rm div\,}\mathbf{B}=0,\quad {\rm div\,}\mathbf{E}=0.
\label{constraints}
\end{eqnarray}
We express the current $\mathbf{j}$ of charged particles in terms of the generalized conductivity $\sigma$
\begin{equation}
\label{current-j}
\mathbf{j}=\frac{1}{a}\sigma \mathbf{E}.
\end{equation}

The electric and magnetic fields are vector quantities and, in addition to the absolute value, are characterized by the direction. Since these fields are generated from quantum fluctuations, they are chaotically oriented. Operating with such vector objects 
is inconvenient. Even more crucially, solving the system of equations (\ref{Friedmann})--(\ref{constraints}) requires determining the 
dependence of electric and magnetic fields on spatial coordinates that makes numerical calculations very demanding. Therefore, we generalize our 
approach in Ref.\cite{Sobol:2018} to the case of the pseudoscalar inflation model and introduce the following expectation 
values of scalar quadratic functions of electric and magnetic fields:
\begin{eqnarray}
\label{EE}
\mathscr{E}^{(n)}&=&\frac{1}{a^{n}}\langle \mathbf{E}\cdot {\rm rot}^{n}\mathbf{E}\rangle,\\
\label{EB}
\mathscr{G}^{(n)}&=&-\frac{1}{a^{n}}\langle \mathbf{E}\cdot {\rm rot}^{n}\mathbf{B}\rangle,\\
\label{BB}
\mathscr{B}^{(n)}&=&\frac{1}{a^{n}}\langle \mathbf{B}\cdot {\rm rot}^{n}\mathbf{B}\rangle.
\end{eqnarray}  
where $n=0,1,2,...$. Note that we consider quadratic correlators of electric and magnetic fields and spatial derivatives of the 
latter through curls because the Maxwell equations (\ref{Maxwell-2}) and (\ref{Maxwell-Bianchi-2}) contain spatial derivatives of 
electric and magnetic fields only in the form of curls. As we will see below, the functions with $n > 1$ are considered because the 
evolution equations for the expectation values with $n=0$ and $n=1$ do not form a closed system. Finally, we would like to note that 
the position of curl acting on the first or the second multiplier does not matter in view of the identity
\begin{equation*}
\langle {\rm rot\,}\mathbf{f}\cdot \mathbf{g}\rangle -\langle \mathbf{f}\cdot {\rm rot\,}\mathbf{g}\rangle={\rm div\,}\langle[\mathbf{f}
\times\mathbf{g}]\rangle =0
\end{equation*}
because the expectation value $\langle[\mathbf{f}\times\mathbf{g}]\rangle$ does not depend on the spatial coordinates.

In fact, functions (\ref{EE})--(\ref{BB}) for $n=0$ are related to the well-known characteristics of the EM field.
Indeed, $\mathscr{E}^{(0)}/2=\rho_{E}$ and $\mathscr{B}^{(0)}/2=\rho_{B}$ are the electric and magnetic field energy densities, 
respectively. Further, $\mathscr{G}^{(0)}=-\langle \mathbf{E}\cdot\mathbf{B}\rangle$ is an important quantity which depends on 
helical properties of the EM field and determines its backreaction on the inflaton.
Moreover, the quantity $\mathscr{B}^{(-1)}$ defines the magnetic helicity [we treat the quantity
${\rm rot}^{-1}\mathbf{B}$ in Eq.~(\ref{BB}) for $n=-1$ simply as $a^{-2}\mathbf{A}$ in accord with Eq.~(\ref{fields-E-and-B})]
\begin{equation}
\mathscr{B}^{(-1)}\equiv \mathcal{H}=\frac{1}{a^{3}}\left<\mathbf{A}\cdot {\rm rot\,}\mathbf{A}\right>=\frac{1}{a}\left<\mathbf{A}
\cdot \mathbf{B}\right>,
\end{equation}
which is crucial for the postinflationary evolution of the magnetic field.

The Maxwell equations (\ref{Maxwell-2}) and (\ref{Maxwell-Bianchi-2}) relate the time derivatives of the electric and magnetic fields 
to their spatial derivatives (curls). Therefore, the evolution equation for any function (\ref{EE})--(\ref{BB}) always contains
a function with one additional spatial curl. This results in an infinite chain of equations for the functions of different 
order. Indeed, by using Eqs.~(\ref{Maxwell-2}) and (\ref{Maxwell-Bianchi-2}), we find
\begin{eqnarray}
\label{eq-E-n}
&& \dot{\mathscr{E}}^{(n)}+[(n+4)H+2\sigma]\mathscr{E}^{(n)}-2I'(\phi)\dot{\phi}\mathscr{G}^{(n)}+2\mathscr{G}^{(n+1)}
=\big[\dot{\mathscr{E}}^{(n)}\big]_{b},\\
\label{eq-G-n}
&&\dot{\mathscr{G}}^{(n)}+[(n+4)H+\sigma]\mathscr{G}^{(n)}-\mathscr{E}^{(n+1)}+\mathscr{B}^{(n+1)}-I'(\phi)\dot{\phi}\mathscr{B}^{(n)}
=\big[\dot{\mathscr{G}}^{(n)}\big]_{b},\\
\label{eq-B-n}
&&\dot{\mathscr{B}}^{(n)}+(n+4)H\mathscr{B}^{(n)}-2\mathscr{G}^{(n+1)}=\big[\dot{\mathscr{B}}^{(n)}\big]_{b}.
\end{eqnarray}
We introduced also extra terms on the right-hand side of the above equations. The reason for their appearance is the following. Electromagnetic
fields are generated from quantum fluctuations and can be treated classically only after the corresponding modes exit the horizon \cite{Lyth:2008} and undergo the tachyonic instability [see the corresponding discussion after Eq.~(\ref{BB-spectrum})]. Therefore, the number of modes, which undergo enhancement, changes as the inflation goes on.
This leads to additional (boundary) terms in the equations of motion (\ref{eq-E-n})--(\ref{eq-B-n}).

Adding Eqs.~(\ref{eq-E-n}) and (\ref{eq-B-n}) for $n=0$ and dividing them by 2, we obtain the equation governing the evolution of the 
EM energy density
\begin{equation}
\label{eq-EM-energy}
\dot{\rho}_{\rm EM}+4H\rho_{\rm EM}+2\sigma \rho_{E}=I'(\phi)\dot{\phi}\mathscr{G}^{(0)}+[\dot{\rho}_{\rm EM}]_{b}.
\end{equation}
The term $2\sigma \rho_{E}$ on the left-hand side of Eq.~(\ref{eq-EM-energy}) describes the dissipation of the EM energy density due 
to the Schwinger effect. It will be considered in more detail in Sec.~\ref{sec-Schwinger}. The first term on the right-hand side corresponds to 
the energy density transfer from the inflaton to the EM field due to the axial 
coupling between them. Obviously, the energy transfer term enters the equation for the inflaton energy density evolution with the
opposite sign [this equation can be obtained from Eq.~(\ref{KGF-2}) multiplying it by $\dot{\phi}$]
\begin{equation}
\label{eq-inflaton-energy}
\dot{\rho}_{\rm inf}+3H(\rho_{\rm inf}+p_{\rm inf})=-I'(\phi)\dot{\phi}\mathscr{G}^{(0)},
\end{equation}
where $\rho_{\rm inf}=\dot{\phi}^{2}/2+V(\phi)$ and $p_{\rm inf}=\dot{\phi}^{2}/2-V(\phi)$.

In contrast to the case of the kinetic coupling model studied in Ref.~\cite{Sobol:2018}, Eq.~(\ref{eq-EM-energy}) for the 
evolution of the EM energy density and the KGF equation (\ref{KGF-2}) do not form a closed system of equations 
because they contain a new quantity $\mathscr{G}^{(0)}=-\langle \mathbf{E}\cdot\mathbf{B}\rangle$. Equation (\ref{eq-G-n}) for $\mathscr{G}^{(0)}$, in turn, contains the higher-order functions $\mathscr{E}^{(1)}$ and $\mathscr{B}^{(1)}$. Clearly, this generates an infinite 
chain of equations and we need to impose some additional approximations in order to terminate it. Our approximation scheme will be discussed in
Sec.~\ref{sec-numerical}. As to the boundary terms on the right-hand side of Eqs.(\ref{eq-E-n})--(\ref{eq-B-n}), their explicit 
expressions will be considered in Sec.~\ref{sec-quant}.

Before closing this section, let us estimate the value of the EM energy density when the backreaction becomes relevant. For this purpose, we consider Eq.~(\ref{KGF-2}) with the coupling function (\ref{coupling-function}) and compare the potential term $V'(\phi)$ which determines the slow rolling of the inflaton with the term on the right-hand side of the KGF equation. Previous numerical studies of magnetogenesis in pseudoscalar inflation \cite{Fujita:2015,Figueroa:2018} found that the electric and magnetic fields make comparable contributions to the energy density, i.e., $\rho_{E}\sim\rho_{B}$ (this is in contrast to the kinetic coupling model where the electric component strongly dominates over the magnetic one, see, e.g., Refs.~\cite{Martin:2008,Demozzi:2009,Kanno:2009,Vilchinskii:2017,Sobol:2018}). Therefore, we can estimate the scalar product $\mathbf{E}\cdot\mathbf{B}$ as follows:
\begin{equation}
|\mathscr{G}^{(0)}|\sim \sqrt{\mathscr{E}^{(0)} \mathscr{B}^{(0)}}=2\sqrt{\rho_{E}\rho_{B}}\sim \rho_{\rm EM}.
\end{equation}
Consequently, the backreaction becomes important when the EM energy density reaches the value
\begin{equation}
\label{estimate-BR}
\rho_{\rm EM}\sim \frac{M_{p}}{\beta}|V'(\phi)|\sim \frac{\sqrt{2\epsilon}}{\beta} \rho_{\rm inf},
\end{equation}
where 
\begin{equation}
\label{slow-roll-epsilon}
\epsilon=\frac{M_{p}^{2}}{2} \left(\frac{V'_{\phi}}{V}\right)^{2}
\end{equation}
is the slow-roll parameter and we used the fact that $\rho_{\rm inf}\approx V(\phi)$ in the slow-roll regime $\epsilon\ll 1$. Since the typical value of the coupling constant, for which the backreaction can take place during inflation, equals $\beta\sim 10$ (see, e.g., Refs.~\cite{Fujita:2015,Figueroa:2018}), we conclude that the corresponding EM energy density is a few orders of magnitude less than that of the inflaton. This means that generated EM fields affect primarily only the evolution of the inflaton field while contribute to the Friedmann equation only indirectly, through the inflaton. It is worth noting that the same picture holds also for the kinetic coupling model \cite{Sobol:2018,Kanno:2009}.

\section{Boundary terms}
\label{sec-quant}

In this section, we determine the boundary terms for the Maxwell equations (\ref{eq-E-n})--(\ref{eq-B-n}), which describe the quantum-to-classical 
transition for the modes crossing the horizon. For this, we should consider the quantized EM field. There is, however, one complication 
connected with the pair production of charged particles by the classical electric field due to the Schwinger effect. The current of these 
particles backreacts on the EM field evolution making it, in general, nonlinear.
However, in order to define the boundary terms, we will neglect the Schwinger effect. This is possible because the boundary terms are 
important during the early stages of inflation when the EM field is weak and each new mode crossing the horizon makes a notable 
contribution. Later, when the EM field is strong and a lot of modes are outside the horizon, the contribution due to
new modes crossing the horizon is negligible. Vice versa, the Schwinger effect is negligible at the beginning of inflation and becomes 
important only close to the end of inflation when the generated electric field becomes strong. Therefore, the boundary term could be derived 
neglecting the Schwinger effect and the Schwinger current can be added phenomenologically in the final system of equations.

In the Coulomb gauge, the EM vector-potential operator can be decomposed over a set of creation and annihilation operators 
of two transverse polarizations
\begin{equation}
\label{quant-operator}
\hat{\mathbf{A}}(t,\mathbf{x})=\int\frac{d^{3}\mathbf{k}}{(2\pi)^{3/2}}\!\!\sum_{\lambda=\pm}\left\{ \boldsymbol{\varepsilon}_{\lambda}(\mathbf{k})\hat{b}_{\lambda,\mathbf{k}}A_{\lambda}(t,\mathbf{k})e^{i\mathbf{k}\cdot\mathbf{x}}+\boldsymbol{\varepsilon}^{*}_{\lambda}(\mathbf{k})\hat{b}^{\dagger}_{\lambda,\mathbf{k}}A^{*}_{\lambda}(t,\mathbf{k})e^{-i\mathbf{k}\cdot\mathbf{x}}\right\},
\end{equation}
where $\boldsymbol{\varepsilon}_{\lambda}(\mathbf{k})$ are two circular polarization 3-vectors, which satisfy the following 
conditions:
\begin{equation}
\mathbf{k}\cdot\boldsymbol{\varepsilon}_{\lambda}(\mathbf{k})=0,\quad \boldsymbol{\varepsilon}^{*}_{\lambda}(\mathbf{k})=\boldsymbol{\varepsilon}_{-\lambda}(\mathbf{k}), \quad [i\mathbf{k}\times\boldsymbol{\varepsilon}_{\lambda}(\mathbf{k})]=\lambda k \boldsymbol{\varepsilon}_{\lambda}(\mathbf{k}), \quad \boldsymbol{\varepsilon}^{*}_{\lambda}(\mathbf{k})\cdot\boldsymbol{\varepsilon}_{\lambda'}(\mathbf{k})=\delta_{\lambda\lambda'}. 
\end{equation}
The creation and annihilation operators have the standard commutation relations
\begin{equation}
[\hat{b}_{\lambda,\mathbf{k}},\,\hat{b}^{\dagger}_{\lambda',\mathbf{k}'}]=\delta_{\lambda\lambda'}\delta^{(3)}(\mathbf{k}-\mathbf{k}').
\end{equation}

Substituting decomposition (\ref{quant-operator}) into Eq.~(\ref{Maxwell-2}) we find the following equation governing the evolution of the mode function:
\begin{equation}
\label{eq-mode-physical}
\ddot{A}_{\lambda}(t,\mathbf{k})+H\dot{A}_{\lambda}(t,\mathbf{k})+\left[\frac{k^{2}}{a^{2}}-\lambda\frac{k}{a}2\xi H\right]A_{\lambda}(t,\mathbf{k})=0
\end{equation}
or in conformal time $\eta=\int^{t}dt'/a(t')$,
\begin{equation}
\label{eq-mode-conformal}
A''_{\lambda}(\eta,\mathbf{k})+\left[k^{2}-2\lambda k \xi a H\right]A_{\lambda}(\eta,\mathbf{k})=0,
\end{equation}
where we introduced the parameter
\begin{equation}
\xi=\frac{1}{2}\frac{I'(\phi)\dot{\phi}}{H}=\frac{\dot{I}}{2H}=\frac{\beta\dot{\phi}}{2M_{p}H}.
\label{parameter-xi}
\end{equation}
In order to define the unique solution of Eqs.~(\ref{eq-mode-physical}) or (\ref{eq-mode-conformal}), we chose the initial conditions for the 
mode function as those in the Bunch-Davies vacuum \cite{Bunch:1978} (see the corresponding discussion in Sec.\ref{sec-initial})
\begin{equation}
\label{Bunch-Davies-vacuum}
A_{\lambda}(\eta,k)=\frac{1}{\sqrt{2k}}e^{-ik\eta}, \quad k\eta\to-\infty.
\end{equation}

Inspecting Eq.~(\ref{eq-mode-physical}), we note that the tachyonic instability takes place for the modes with helicity
$\lambda={\rm sign}(\xi)$ and $k/(aH)\lesssim 2|\xi|$ and one can expect the exponential amplification of these modes. As was mentioned above, 
the number of such modes changes in time.
To derive the explicit expressions of the boundary terms we need to express Eqs.~(\ref{EE})--(\ref{BB}) in terms of the Fourier modes 
of the EM field. Using decomposition (\ref{quant-operator}) and taking into account that only one polarization with $\lambda={\rm sign}(\xi)$ 
undergoes enhancement, we obtain
\begin{eqnarray}
\label{EE-spectrum}
\mathscr{E}^{(n)}&=&\int_{0}^{k_{c}}\frac{dk}{k}\lambda^{n}\frac{k^{n+3}}{2\pi^{2}a^{n+2}}\left|\frac{d}{dt}A_{\lambda}(t,k)\right|^{2},\\  
\label{EB-spectrum}
\mathscr{G}^{(n)}&=&\int_{0}^{k_{c}}\frac{dk}{k}\lambda^{n+1}\frac{ k^{n+4}}{4\pi^{2}a^{n+3}}\frac{d}{dt}\left|{A}_{\lambda}(t,k)\right|^{2},  \\
\label{BB-spectrum}
\mathscr{B}^{(n)}&=&\int_{0}^{k_{c}}\frac{dk}{k}\lambda^{n}\frac{k^{n+5}}{2\pi^{2}a^{n+4}}\left|{A}_{\lambda}(t,k)\right|^{2}, 
\end{eqnarray}
where $k_{c}=2|\xi| a H$ is the critical momentum for which the instability occurs.
The upper integration limit in Eqs.~(\ref{EE-spectrum})--(\ref{BB-spectrum}) is defined by $k_c$ because subhorizon modes have an 
oscillatory behavior and should be regarded as quantum fluctuations. Therefore, such modes do not contribute to the classical observables 
(\ref{EE})--(\ref{BB}) and are excluded from the integration.
Throughout the paper, we refer to a mode as subhorizon if its momentum is larger than the critical one $k>k_{c}$ [its wavelength is smaller than the effective horizon size $1/(2|\xi|H)$] and superhorizon in the opposite case.

In order to derive the boundary term, we note that Eq.~(\ref{eq-mode-conformal}) by using the replacements
$$
\eta \leftrightarrow x,\quad k^{2}\leftrightarrow\frac{2mE}{\hbar^{2}},\quad 2a(\eta)H |\xi| k\leftrightarrow\frac{2mV(x)}{\hbar^{2}},\quad
A_{\lambda} \leftrightarrow \Psi,
$$
can be rewritten in the form of the Schr\"{o}dinger equation
\begin{equation}
\frac{d^{2}\Psi}{dx^{2}}+\frac{2m}{\hbar^{2}}\left(E-V(x)\right)=0.
\end{equation}
For modes with momentum $k\ll k_{c}$, one can apply the Wentzel-Kramers-Brillouin (WKB) approximation and find the following approximate 
solution of Eq.~(\ref{eq-mode-conformal}) with the boundary conditions (\ref{Bunch-Davies-vacuum}):
\begin{equation}
\label{solution-appr}
A_{\lambda}(k,\eta)\simeq\frac{1}{\sqrt{2p/\hbar}}\exp\left[-\frac{i}{\hbar}\int^{\eta}p(\eta')d\eta'\right],
\end{equation}
where $\frac{p}{\hbar}=\sqrt{k^{2}-2aH|\xi|k}$.
This solution is applicable when
\begin{equation}
\left|\frac{d}{d\eta}\frac{\hbar}{p}\right|\approx \frac{\left(\frac{aH}{k}\right)^{2}|\xi|}{\left(1-2|\xi|\frac{aH}{k}\right)^{3/2}}\ll 1.
\end{equation}
Here only the scale factor $a$ has been differentiated as the fastest varying quantity.

By using $k=2|\xi|aH (1+\delta)$, we obtain that the semiclassical solution (\ref{solution-appr}) is applicable for
\begin{equation}
\label{epsilon}
\delta^{3/2}(1+\delta)^{1/2}\gg (4|\xi|)^{-1}.
\end{equation}
Clearly, the semiclassical approximation fails when the left-hand and right-hand sides in the above inequality are approximately equal.
A solution of the corresponding equation can be easily found in the two limiting cases $|\xi|\gg 1$ ($\delta_{c}\simeq [4|\xi|]^{-2/3}$) and $|\xi|\ll 1$ ($\delta_{c}\simeq [4|\xi|]^{-1/2}$)
and then fit to the exact solution with accuracy better than 3\% for all values of $\xi$ by the following function:
\begin{equation}
\label{eps-approx}
\delta_{c}\simeq (4|\xi|)^{-1/2}\left[1+(4|\xi|)^{2/3}\right]^{-1/4}.
\end{equation}

As usual, the WKB method is inapplicable near the ``classical turning point'' $k=k_{c}$. Still we can approximate the value of $A_{\lambda}(t,k_{c})$ by its value
at the boundary of validity of the semiclassical approximation at a slightly larger momentum $k_{c}(1+\delta_{c})$.
Using Eqs.~(\ref{solution-appr}), (\ref{epsilon}), and (\ref{eps-approx}), we have
\begin{eqnarray}
\left|{A}_{\lambda}(t,k_{c})\right|^{2}&\approx&\frac{1}{2k_c}[\delta_{c}(1+\delta_{c})]^{-1/2}=\frac{1}{2k_c} Q(|\xi|)  ,\\
\left|\dot{A}_{\lambda}(t,k_{c})\right|^{2}&\approx&\frac{1}{2k_c}\frac{k_c^{2}}{a^{2}}[\delta_{c}(1+\delta_{c})]^{1/2}= \frac{1}{2k_c} \frac{k_c^{2}}{a^{2}} \frac{1}{Q(|\xi|)}, \\ \frac{d}{dt}\left|{A}_{\lambda}(t,k_{c})\right|^{2}&\approx&\frac{1}{2k_c}\frac{k_c}{a}\frac{1}{4|\xi|}\frac{1}{\delta_{c}^{3/2}(1+\delta_{c})^{1/2}}= \frac{1}{2k_c}\frac{k_c}{a},
\end{eqnarray}
where $Q(|\xi|)=4|\xi|\delta_{c}\approx (4|\xi|)^{1/2}\left[1+(4|\xi|)^{2/3}\right]^{-1/4}$.
Then the boundary terms equal
\begin{eqnarray}
\big[\dot{\mathscr{E}}^{(n)}\big]_{b}&=&\lambda^{n}\frac{k_c^{n+3}}{2\pi^{2}a^{n+2}}\left|\dot{A}_{\lambda}(t,k_{c})\right|^{2} \frac{d \ln k_{c}}{dt}\simeq \frac{H}{4\pi^{2}}(2H\xi)^{n+4} \frac{1}{Q(|\xi|)},\label{bound-E}\\
\big[\dot{\mathscr{G}}^{(n)}\big]_{b}&=&\lambda^{n+1}\frac{k_c^{n+4}}{4\pi^{2}a^{n+3}}\frac{d}{dt}\left|A_{\lambda}(t,k_{c})\right|^{2} \frac{d \ln k_{c}}{dt}\simeq \frac{\lambda H}{8\pi^{2}}(2H\xi)^{n+4},\label{bound-G}\\
\big[\dot{\mathscr{B}}^{(n)}\big]_{b}&=&\lambda^{n}\frac{k_c^{n+5}}{2\pi^{2}a^{n+4}}\left|A_{\lambda}(t,k_{c})\right|^{2} \frac{d \ln k_{c}}{dt}\simeq \frac{H}{4\pi^{2}}(2H\xi)^{n+4} Q(|\xi|).\label{bound-B}
\end{eqnarray} 

Finally, let us comment on the magnetic helicity and its connection with the correlation length of the magnetic field. 
The energy density and helicity of the magnetic field are given by \cite{Durrer:2013,Subramanian:2016}
\begin{eqnarray}
\rho_{B}&=&\int_{0}^{k_{c}}\frac{dk}{k}\frac{k^{5}}{4\pi^{2}a^{4}}\left[\left|{A}_{+}(t,k)\right|^{2}+\left|{A}_{-}(t,k)\right|^{2}\right],\\
\mathcal{H}&=&\int_{0}^{k_{c}}\frac{dk}{k}\frac{k^{4}}{2\pi^{2}a^{3}}\left[\left|{A}_{+}(t,k)\right|^{2}-\left|{A}_{-}(t,k)\right|^{2}\right] \label{helicity}
\end{eqnarray}
and its correlation length equals
\begin{equation}
\lambda_{B}=\left<\frac{2\pi a}{k}\right>_{k}=\frac{1}{\rho_{B}}\int_{0}^{k_{c}}\frac{dk}{k}\frac{2\pi a}{k}\frac{d\rho_{B}}{d\ln k}=\frac{\pi}{\rho_{B}}\int_{0}^{k_{c}}\frac{dk}{k}\frac{k^{4}}{2\pi^{2}a^{3}}\left[\left|{A}_{+}(t,k)\right|^{2}+\left|{A}_{-}(t,k)\right|^{2}\right].\label{corr-length}
\end{equation}
Comparing Eqs.~(\ref{helicity}) and (\ref{corr-length}), we derive the so-called \textit{realizability condition}:
\begin{equation}
|\mathcal{H}|\leq \frac{\lambda_{B}\rho_{B}}{\pi}.
\end{equation}
This inequality is correct for any configuration of the magnetic field. However, in the case of maximally helical magnetic field, 
where only one polarization is enhanced, it approaches the equality
\begin{equation}
\label{corr-length-B}
\lambda_{B}=\frac{\pi |\mathcal{H}|}{\rho_{B}}=\frac{2\pi |\mathscr{B}^{(-1)}|}{\mathscr{B}^{(0)}}.
\end{equation}
which can be used to determine the correlation length of the magnetic field.

\section{Initial conditions}
\label{sec-initial}

It is well known that in order to resolve several cosmological problems, such as the flatness and horizon problems, one needs
at least 50--60 $e$-foldings of inflation. Moreover, the primordial power spectrum inspected by the CMB observations is generated also at 50--60 $e$-foldings before the end of inflation. Therefore, although inflation can last more $e$-foldings, it is natural in numerical analysis to take the starting point of the simulation at 60 $e$-foldings before the end of inflation. Then, in order to find a solution to Eqs.~(\ref{eq-E-n})--(\ref{eq-B-n}), one should specify the initial conditions for the corresponding functions.

As we discussed above, modes deep inside the horizon oscillate and correspond to solution (\ref{Bunch-Davies-vacuum}) in the Bunch-Davies vacuum. In this regime, they represent quantum fluctuations and do not contribute to the classical observables. Later due to the Universe expansion, the physical momentum of a given mode decreases and the square of the effective frequency changes its sign causing a tachyonic instability. The corresponding mode function stops to oscillate and can be interpreted as a classical field. Therefore, in order to impose the initial conditions, we should take into account all modes which underwent the tachyonic instability before the initial moment of time.

Let us consider how the parameter $\xi$ evolves during the slow-roll inflation in the absence of the backreaction of generated EM fields. Using 
the KGF equation (\ref{KGF-2}) for the inflaton and the Friedmann equation (\ref{Friedmann}) in the slow-roll regime, we
find
\begin{equation}
\label{xi-small}
|\xi|=\frac{I'(\phi)|\dot{\phi}|}{2H}\simeq\frac{I'(\phi)}{2}\frac{|V'_{\phi}|}{3H^2}\simeq  \frac{I'(\phi) M_{p}}{\sqrt{2}}\sqrt{\epsilon_{1}},
\end{equation}
where $\epsilon_{1}$ is the first Hubble-flow function which coincides with the slow-roll parameter $\epsilon$ [see Eq.~(\ref{slow-roll-epsilon})] in the slow-roll regime \cite{Martin:2014}.
The rate of time evolution of $\xi$ is given by
\begin{equation}
\label{rate-of-change}
\frac{1}{H}\left|\frac{\dot{\xi}}{\xi}\right|=M_{p}\frac{I''(\phi)}{I'(\phi)}\sqrt{2\epsilon_{1}}+\frac{1}{2}\epsilon_{2}\ll 1,
\end{equation}
where $\epsilon_{2}$ is the second Hubble-flow function from an infinite sequence of such quantities \cite{Martin:2014}
\begin{equation}
\label{Hubble-flow}
\epsilon_{n+1}=\frac{d \ln \epsilon_{n}}{d \ln a}, \quad \epsilon_{0}=\frac{H_{i}}{H},
\end{equation} 
and $H_{i}$ is the initial value of the Hubble parameter, which is used as a reference point.

For the axionlike coupling (\ref{coupling-function}), only the second term in Eq.~(\ref{rate-of-change}) remains. Then
we conclude that during inflation $\xi$ changes adiabatically slowly and can be considered as a constant for a few 
Hubble times. On the other hand, over long periods of time it slightly changes and its value is determined by the current values of 
$\dot{\phi}$ and $H$. In other words, it could be taken as a constant parameter in the Maxwell equations, therefore, the generation 
of EM fields is governed only by the value of $\xi$ at a given moment of time. 
We can also approximate the space-time metric by the de Sitter one assuming that the scale factor grows strictly exponentially
\begin{equation}
\label{de-Sitter}
a(t)=e^{Ht}=-\frac{1}{H\eta}.
\end{equation}
Clearly, this approximation is valid with high accuracy in the slow-roll regime far from the end of inflation.
Then Eq.~(\ref{eq-mode-conformal}) takes the form
\begin{equation}
\label{mode-equation-xi}
A''_{\lambda}(\eta, \mathbf{k})+\left[k^{2}+\lambda\frac{2k\xi}{\eta}\right]A_{\lambda}(\eta, \mathbf{k})=0.
\end{equation}
Since the conformal time during the inflation stage has the negative sign, the second term in 
brackets in Eq.~(\ref{mode-equation-xi}) for $\lambda={\rm sign}(\xi)$ is negative and, for a certain range of momenta $k/(aH)<2|\xi|$, it becomes larger than the first 
term and causes the tachyonic instability for the corresponding modes. This instability is absent for modes with another 
polarization, therefore, generated EM fields are helical. For $|\xi|\gtrsim 1$, the amplification is very effective and the 
corresponding fields can be considered as maximally helical, i.e., one can simply neglect the second circular polarization as we do below.

According to Ref.~\cite{Anber:2010}, the solution of Eq.~(\ref{mode-equation-xi}) can be found in terms of Coulomb wave functions \cite{Luke:book}. However, it 
is more convenient to use the approximate expression for the growing mode for $k/(aH)<2|\xi|$ found in Ref.~\cite{Durrer:2011}
\begin{equation}
\label{solution-adiabatic-approximation}
A_{\lambda}=\frac{1}{\sqrt{2k}}W(|\xi|)z K_{1}(z), \quad \dot{A}_{\lambda}=\frac{1}{\sqrt{2k}}\frac{H}{2}W(|\xi|)z^{2} K_{0}(z),
\end{equation}
where $W(|\xi|)=[e^{\pi |\xi|}\,{\rm sh\,}\pi |\xi|/(\pi |\xi|)]^{1/2}$, $z=2\sqrt{2|\xi|}\sqrt{k/(aH)}$, and $K_{\nu}(z)$ is the Macdonald function.
Substituting these expressions into Eqs.~(\ref{EE-spectrum})--(\ref{BB-spectrum}), we obtain the initial conditions for the corresponding expectation values as functions of the parameter $\xi_{0}$ at the initial moment of time:
\begin{equation}
\label{EE-apr-2}
\mathscr{E}^{(n)}(0)=\lambda^{n}\frac{H^{n+4}W^{2}(|\xi_{0}|)}{2^{3n+9}\pi^{2}|\xi_{0}|^{n+2}}\int_{0}^{4|\xi_{0}|}z^{2n+7}K_{0}^{2}(z)dz,
\end{equation}
\begin{equation}
\label{EB-apr-2}
\mathscr{G}^{(n)}(0)=\lambda^{n+1}\frac{H^{n+4}W^{2}(|\xi_{0}|)}{2^{3n+11}\pi^{2}|\xi_{0}|^{n+3}}\int_{0}^{4|\xi_{0}|}z^{2n+8}K_{0}(z)K_{1}(z)dz,
\end{equation}
\begin{equation}
\label{BB-apr-2}
\mathscr{B}^{(n)}(0)=\lambda^{n}\frac{H^{n+4}W^{2}(|\xi_{0}|)}{2^{3n+13}\pi^{2}|\xi_{0}|^{n+4}}\int_{0}^{4|\xi_{0}|}z^{2n+9}K_{1}^{2}(z)dz,
\end{equation}
These integrals can be expressed in terms of the Meijer $G$ function and the corresponding results are listed in Appendix~\ref{App-math} [see 
Eqs.~(\ref{EE-apr-4})--(\ref{EB-apr-4})] together with their asymptotic expressions in the two limiting cases $|\xi_{0}|\ll 1$ and $|\xi_{0}|\gg 1$.

\section{Schwinger effect}
\label{sec-Schwinger}

In this section, we determine the Schwinger conductivity $\sigma$ and the energy density of charged particles $\rho_{\chi}$ created via the Schwinger process.
For the Schwinger current, we will use the expressions derived in the minimal subtraction scheme in Refs.~\cite{Kobayashi:2014,Hayashinaka:2016a}. Since the general expressions are rather cumbersome, it is more convenient to use 
their asymptotics governed by the parameters
\begin{equation}
M=\frac{m}{H}, \quad L=\frac{eE}{H^{2}}=e\sqrt{\frac{\mathscr{E}^{(0)}}{H^{4}}},
\end{equation}
where $m$ and $e$ are the mass and electric charge of created particles, respectively.

The typical value of the Hubble parameter, which can be fixed from the observations of the amplitude of primordial perturbations
\cite{Planck:2018-infl}, is $H\sim 10^{-5}\,M_{p}\sim 10^{13}\,{\rm GeV}$. Since the lightest charged spin-$1/2$ particle, the 
electron, has a mass $m\sim 10^{-3}\,{\rm GeV}$ and the lightest charged scalar particle, the pion, has a mass $m\sim 0.1\,{\rm GeV}$,
we are interested in the small mass limit $M\ll 1$.

At the beginning of inflation, the parameter $\xi$ is usually small because it is proportional to the square root of the slow-roll parameter 
$\epsilon$, see Eq.~(\ref{xi-small}). Thus, we may use Eq.~(\ref{EE-appr-small}) to estimate the initial value of the electric 
energy density. Then the parameter $L$ equals
\begin{equation}
L_{0}\simeq \frac{4e}{\pi} |\xi_{0}^{3}\ln(2b|\xi_{0}|)|\ll 1,
\end{equation}
where $b=\exp(\gamma_{E})$ and $\gamma_{E}=0.577...$ is the Euler-Mascheroni constant.
For a general model of inflation consistent with the CMB observations, $\epsilon$ at 50--60 $e$-foldings before the end of inflation can be expressed in terms of the tensor-to-scalar power ratio as $\epsilon=r/16\lesssim 10^{-3}$ (according to the latest CMB observations by Planck Collaboration and BICEP-
Keck Array \cite{Planck:2018-infl}, the upper limit on $r$ is $r<0.064$). Then, for the coupling constant $\beta\sim 10$, 
we have $|\xi_{0}|\sim 0.1$ and $L_{0}\sim 10^{-4}$.

On the other hand, close to the end of inflation and for the sufficiently large couplings that can cause the backreaction, the electric energy 
density is given by
$\mathscr{E}^{(0)}/2=\rho_{E}\sim \sqrt{\epsilon}\beta^{-1}\rho_{\rm tot}=3\sqrt{\epsilon}\beta^{-1}H^{2}M_{p}^{2}$ 
[see estimate (\ref{estimate-BR})]. Then
$L\sim e \epsilon^{1/4}\beta^{-1/2} M_{p}/H\sim 10^{2}-10^{3}$ for $H\sim 10^{-5}\,M_{p}$ and
$\beta\sim 10$. Therefore, the values of $L$ can vary in a wide range.

In the weak-field regime, $L\ll 1$, the fermionic Schwinger conductivity has the form \cite{Hayashinaka:2016a}
\begin{equation}
\label{current-fermi}
\sigma_{f}= \frac{e^{2}H}{18\pi^{2}}\left(6 \ln \frac{m}{H}-6\gamma_{E}-1\right)=\frac{2\alpha_{\rm EM}}{9\pi}\left(6 \ln \frac{m}{H}-6\gamma_{E}-1\right) H,
\end{equation}
where $\alpha_{\rm EM}=e^{2}/(4\pi)\approx 1/137$ is the fine structure constant. It is easy to notice that the conductivity is negative in the low mass limit $m\ll H$ and could, in principle, enhance the electric field. Such a possibility was investigated in Ref.~\cite{Stahl:2018} where it was found
that the enhancement is negligible compared to the damping of the field due to the Universe expansion because the Schwinger conductivity is much less by the absolute value than the Hubble parameter for physically relevant particle masses. Moreover, the authors of Ref.~\cite{Banyeres:2018} claim that the negative conductivity is spurious and corresponds to nonlinear corrections to the Maxwell action and the logarithmic running of the coupling constant. In any case, the ratio $\sigma/H$ is always small and the Schwinger effect is negligible in the small-field regime.

For bosons, the conductivity is positive and attains very large values in the low mass regime where it diverges like $1/M^{2}$ for $M\to 0$ \cite{Kobayashi:2014,Hayashinaka:2016b}. In the literature this phenomenon is known as the infrared hyperconductivity. Since such a behavior is observed only in the extremely low field regime when $L\ll M\ll 1$, it is never realized in the problem under consideration because as we showed above $L$ cannot take extremely small values. For $L > M$, the conductivity quickly diminishes to values typical in the fermionic case (\ref{current-fermi}) (see Ref.~\cite{Hayashinaka:2016b}). Thus, the Schwinger effect does not play any role in the weak field regime.

The strong-field expressions ($L\gg 1$) for the Schwinger conductivity are similar for bosons and fermions and have the form
\begin{equation}
\label{schwinger-current-large-field}
\sigma_{s}=\frac{g_{s}}{12\pi^{3}}\frac{e^{3}E}{H}e^{-\frac{\pi m^{2}}{|eE|}}, \quad s={b,\,f},
\end{equation}
where $g_{b}=1$ and $g_{f}=2$ are the number of spin degrees of freedom. Expressing the conductivity in terms of the energy density 
and neglecting the particles mass, we obtain
\begin{equation}
\label{strong-field}
\sigma_{s}=\frac{g_{s}e^{3}}{12\pi^{3}}\frac{\sqrt{\mathscr{E}^{(0)}}}{H}=\frac{\alpha_{\rm EM}g_{s}}{3\pi^{2}} L H.
\end{equation}
Then we conclude that the Schwinger effect is important only in the strong field regime when $L> 3\pi^{2}/\alpha_{\rm EM}\gg 1$.
The corresponding condition for the electric energy density reads as
\begin{equation}
\label{condition-Schwinger}
\rho_{E}>\frac{9\pi^{3}}{8 \alpha_{\rm EM}^{3}}H^{4}= \frac{\pi^{3}}{8 \alpha_{\rm EM}^{3}} \frac{\rho_{\rm tot}^{2}}{M_{p}^{4}}.
\end{equation}
Therefore, the strong-field 
expression (\ref{strong-field}) could be used all time. Although this expression is not correct in the weak field regime, this is irrelevant for the dynamics of EM fields.

Finally, we need also an equation governing the evolution of created particles. It is natural to require that the energy dissipated by EM fields is transferred into created particles.
Then the energy density of produced particles can be described phenomenologically by the following equation:
\begin{equation}
\label{produced-density}
\dot{\rho}_{\chi}+4H\rho_{\chi}=\sigma_{s} \mathscr{E}^{(0)},
\end{equation}
where the term on the right-hand side is exactly the same as in Eq.~(\ref{eq-EM-energy}) for the electric energy density, however, 
has the opposite sign. Since we consider the particle mass to be smaller 
than the Hubble parameter, we treat the created particles as ultrarelativistic that gives the factor $4H$ in the above equation.

\section{Numerical results}
\label{sec-numerical}

Having derived an infinite chain of Eqs.~(\ref{eq-E-n})--(\ref{eq-B-n}), we introduce in this section an approximation scheme which allows us to truncate this chain of equations and then numerically solve it in a certain inflationary model. The crucial physical input that allows us to perform this truncation is the 
assumption that electric and magnetic fields contribute equally to the energy density. This is a numerically well-justified approximation, 
especially for the small values of the coupling constant $\beta\lesssim 10$ (see, e.g., Fig.~6 in Ref.~\cite{Figueroa:2018}). The 
generalization of this approximation scheme to higher order quantities is straightforward. Indeed, we can assume that
$\mathscr{E}^{(n)}=\mathscr{B}^{(n)}$ for a given $n$. This makes it possible to keep only 2 equations in the $n$th order. The first 
is the equation for $\mathscr{G}^{(n)}$, where $\mathscr{E}^{(n+1)}=\mathscr{B}^{(n+1)}$ is assumed and $\mathscr{B}^{(n)}$ is replaced by
$\mathscr{E}^{(n)}$,
\begin{equation}
\dot{\mathscr{G}}^{(n)}+[(n+4)H+\sigma]\mathscr{G}^{(n)}-\dot{I}\mathscr{E}^{(n)}=\big[\dot{\mathscr{G}}^{(n)}\big]_{b}.
\end{equation}
The second equation for $(\mathscr{E}^{(n)}+\mathscr{B}^{(n)})/2\equiv\mathscr{E}^{(n)}$,
\begin{equation}
\dot{\mathscr{E}}^{(n)}+[(n+4)H+\sigma]\mathscr{E}^{(n)}-\dot{I}\mathscr{G}^{(n)}=\frac{1}{2}\left\{ \big[\dot{\mathscr{E}}^{(n)}\big]_{b}
+\big[\dot{\mathscr{B}}^{(n)}\big]_{b}\right\}
\end{equation}
is obtained as the sum of equations (\ref{eq-E-n}) and (\ref{eq-B-n}) divided by 2. Clearly, the two above equations together
with the Friedmann equation (\ref{Friedmann}) for the scale factor, the KGF equation (\ref{KGF-2}) for the inflaton, and the equation for the helicity $\mathcal{H}$ form a
closed system of evolution equations. Thus, in the approximation scheme of order $n$, we deal with $3n+5$ ordinary differential equations.

Below we provide numerical solutions of the system of equations discussed above for the inflationary model with concave and even 
potential. The latter property ensures the parity symmetry is preserved for the pseudoscalar inflaton field. The requirement for the inflaton 
potential to be concave follows from the recent CMB observations \cite{Planck:2018-infl}. The best correspondence with observational data have 
the plateau models. The $T$ model of $\alpha$ attractors with hyperbolic tangent potential
\cite{Kallosh:2013a,Ferrara:2013,Kallosh:2013b}
\begin{equation}
V(\phi)=\Lambda^{4} {\rm th}^{2}\frac{\phi}{\sqrt{6\alpha}M_{p}}
\end{equation}
is a typical representative of the plateau models. Here $\alpha$ is a free parameter. For given values of $\alpha$ and the number of 
$e$-foldings before the end of inflation when the pivot mode crosses the Hubble horizon $N_{\ast}$, one can determine the scalar spectral index 
$n_{s}$ and the tensor-to-scalar power ratio $r$
\begin{eqnarray}
n_{s}&=&1-6\epsilon_{\ast}+2\eta_{\ast}=1-\frac{8N_{\ast}+6\alpha+2\sqrt{3\alpha(3\alpha+4)}}{4N_{\ast}^{2}+2N_{\ast}\sqrt{3\alpha(3\alpha+4)}+3\alpha},\\
r&=&16\epsilon_{\ast}=\frac{48 \alpha}{4N_{\ast}^{2}+2N_{\ast}\sqrt{3\alpha(3\alpha+4)}+3\alpha}.
\end{eqnarray}
Varying $\alpha$ and $N_{\ast}$ one can fill the region on the $(n_s,\, r)$ diagram and compare the predictions of the model with the constraints set by the CMB observations of Planck Collaboration \cite{Planck:2018-infl}. In particular, for $N_{\ast}=60$, we found that the range of $\alpha$ compatible with the 68\% C.L. constraints is $\alpha<20$. For definiteness, we use $\alpha=1$ in the numerical analysis. For this value of the parameter $\alpha$, the potential has the form which is very similar to that of the well-known Starobinsky model, which is another typical plateau model.

Finally, we should determine the strength of the potential which gives the correct value of 
the scalar power spectrum amplitude in Eq.~(\ref{amplitude}) below,
\begin{equation}
\left(\frac{\Lambda}{M_{p}}\right)^{4}=12\pi^{2}A_{s}\frac{6 \alpha}{4 N_{\ast}^{2}+2N_{\ast}\sqrt{3\alpha(3\alpha+4)}+3\alpha}
\frac{4N_{\ast}+3\alpha+\sqrt{3\alpha(3\alpha+4)}}{4N_{\ast}-3\alpha+\sqrt{3\alpha(3\alpha+4)}}.
\end{equation}
Thus, fixing $N_{\ast}$ and $\alpha$, we fully determine the inflaton potential. Further, for a certain value of the coupling constant $\beta$, 
we calculate $\xi_{0}$ and use it to determine the initial conditions given by Eqs.~(\ref{EE-apr-3})--(\ref{BB-apr-3}). For 
definiteness, we consider the case of fermion charged particles created due to the Schwinger effect and then calculate their electric conductivity by using Eq.~(\ref{strong-field}). Then, we apply the $n$th order approximation scheme and compute numerically the time dependence of the inflaton, scale factor, and EM quadratic functions (\ref{EE})--(\ref{BB}). 

\begin{figure}[h!]
	\centering
	\includegraphics[height=5.3cm]{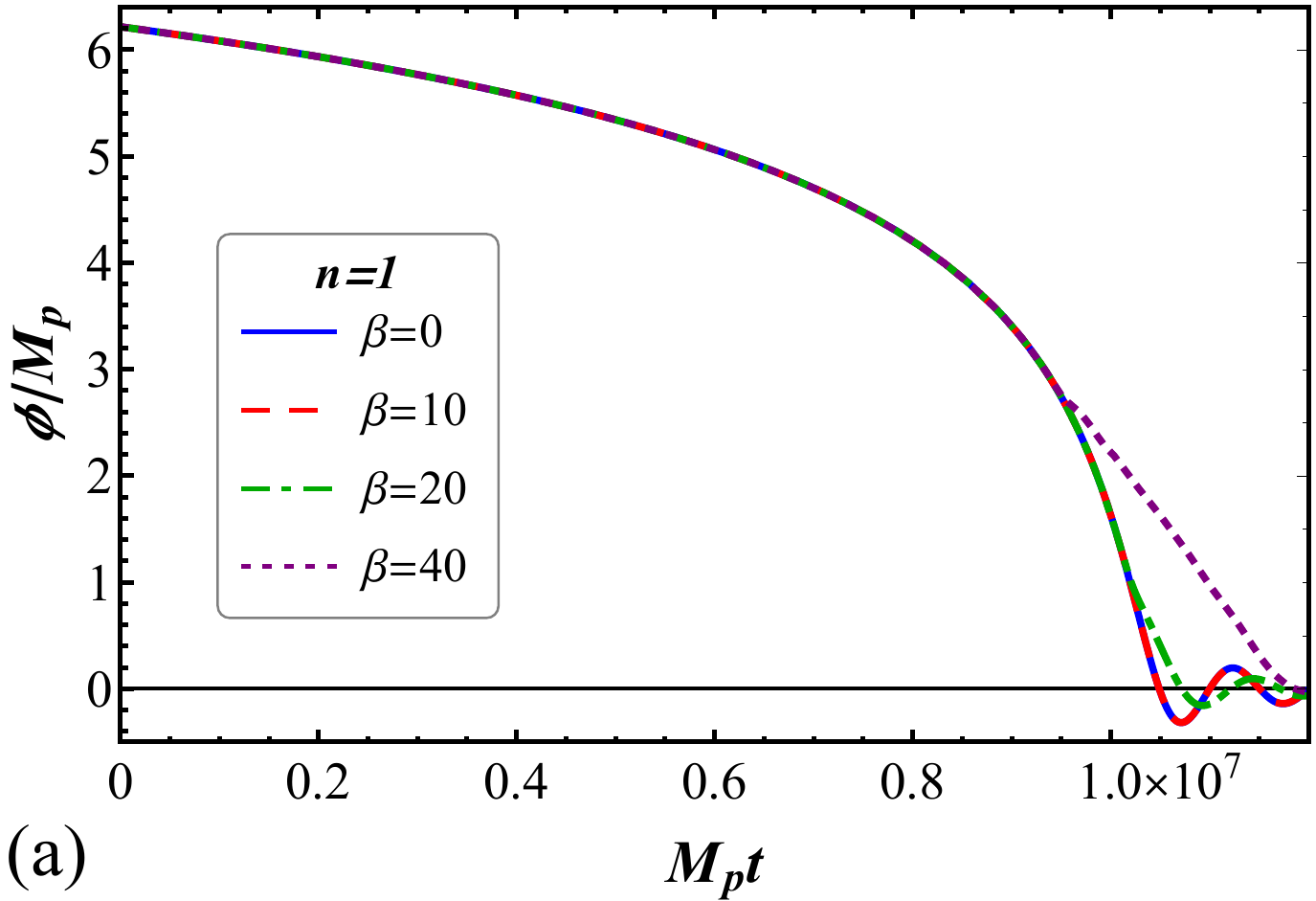} \hfill
	\includegraphics[height=5.2cm]{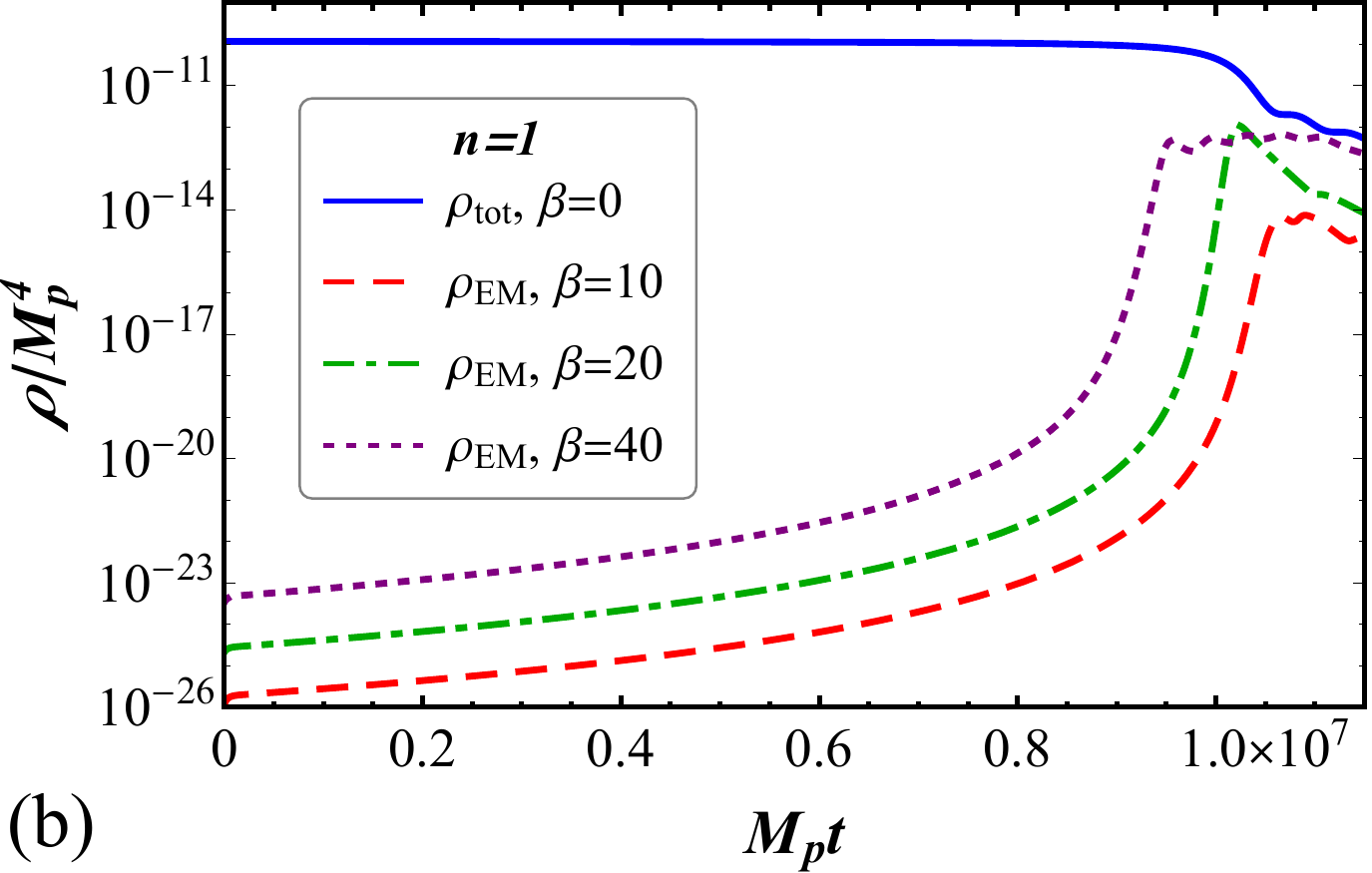}
	\caption{The time dependence of the inflaton field (a) and the EM energy density (b) calculated using the approximation scheme with $n=1$ for three different values of the coupling constant: $\beta=10$ (red dashed line), $\beta=20$ (green dashed-dotted line), and $\beta=40$ (purple dotted line). The blue solid lines show the time dependence of the inflaton field and the total energy density $\rho_{tot}=3H^{2}M_{p}^{2}$ for $\beta=0$.\label{fig-1-comparison}}
\end{figure}

Only relatively small values of the coupling constant $\beta \lesssim 10$ are physically allowed, because for larger values the reheating stage is spoiled and large non-Gaussianities in primordial spectra are generated \cite{Fujita:2015,Figueroa:2018}. Nevertheless, we will apply our method also for larger values of $\beta$ in order to study its applicability and to show more clearly some characteristic features of the magnetogenesis process. Figure~\ref{fig-1-comparison} shows the time dependence of the inflaton field [panel (a)] and the EM energy density [panel (b)] during the inflation stage calculated in the approximation scheme with $n=1$ for three nonzero values of the coupling constant. The curves for the smallest value $\beta=10$ are shown by the red dashed lines. In this case, the inflaton field evolves in the same way as in the absence of the coupling $\beta=0$ (shown by the blue solid line), meaning that the backreaction of the generated EM field on the inflaton evolution is negligible. 

On the other hand, for larger values of the coupling constant $\beta=20$ and $\beta=40$ shown by the green dashed-dotted line and the purple dotted line, respectively, the time dependence of the energy density at a certain moment of time begins to deviate from the unperturbed solution and changes from fast exponential growth to almost constant until the preheating stage when it starts to decay. As we showed in Sec.~\ref{sec-basics}, the critical value of the EM energy density when the backreaction becomes relevant is of the order $\rho_{\rm EM}\sim (\sqrt{2\epsilon}/\beta)\rho_{\rm inf}$. In addition, for larger couplings, the initial values and the boundary terms are larger too leading to the generation of stronger fields. Consequently, the backreaction takes place earlier for larger coupling constant.  These features are easily seen in Fig.~\ref{fig-1-comparison}(b). The inflaton field also changes its behavior in the backreaction regime evolving more slowly and reaching the potential minimum where it oscillates with small amplitude. Slowing down of the inflaton rolling stops the generation of EM fields which attain constant values.

\begin{figure}[h!]
	\centering
  	\includegraphics[height=5.1cm]{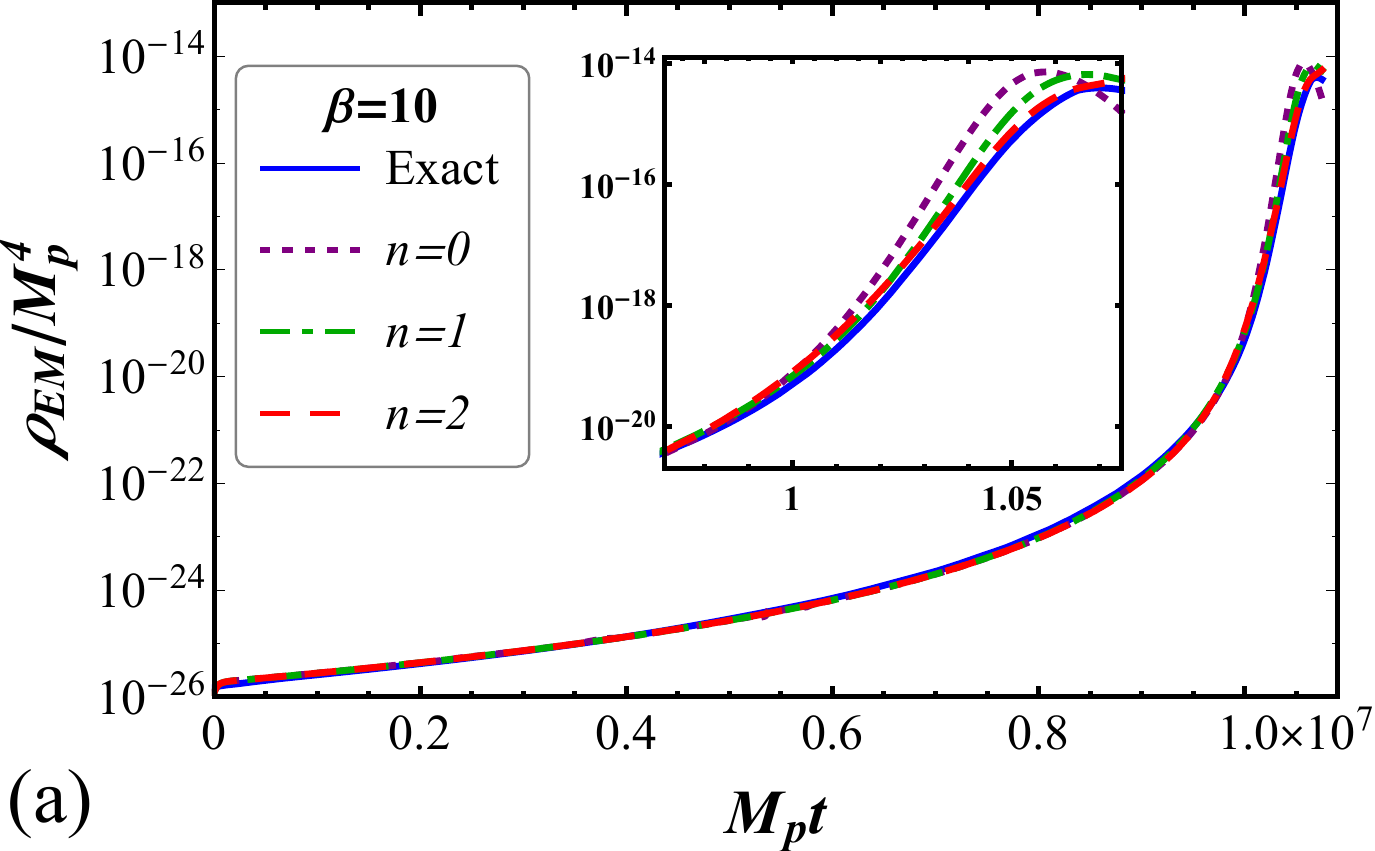} \hfill
	\includegraphics[height=5.1cm]{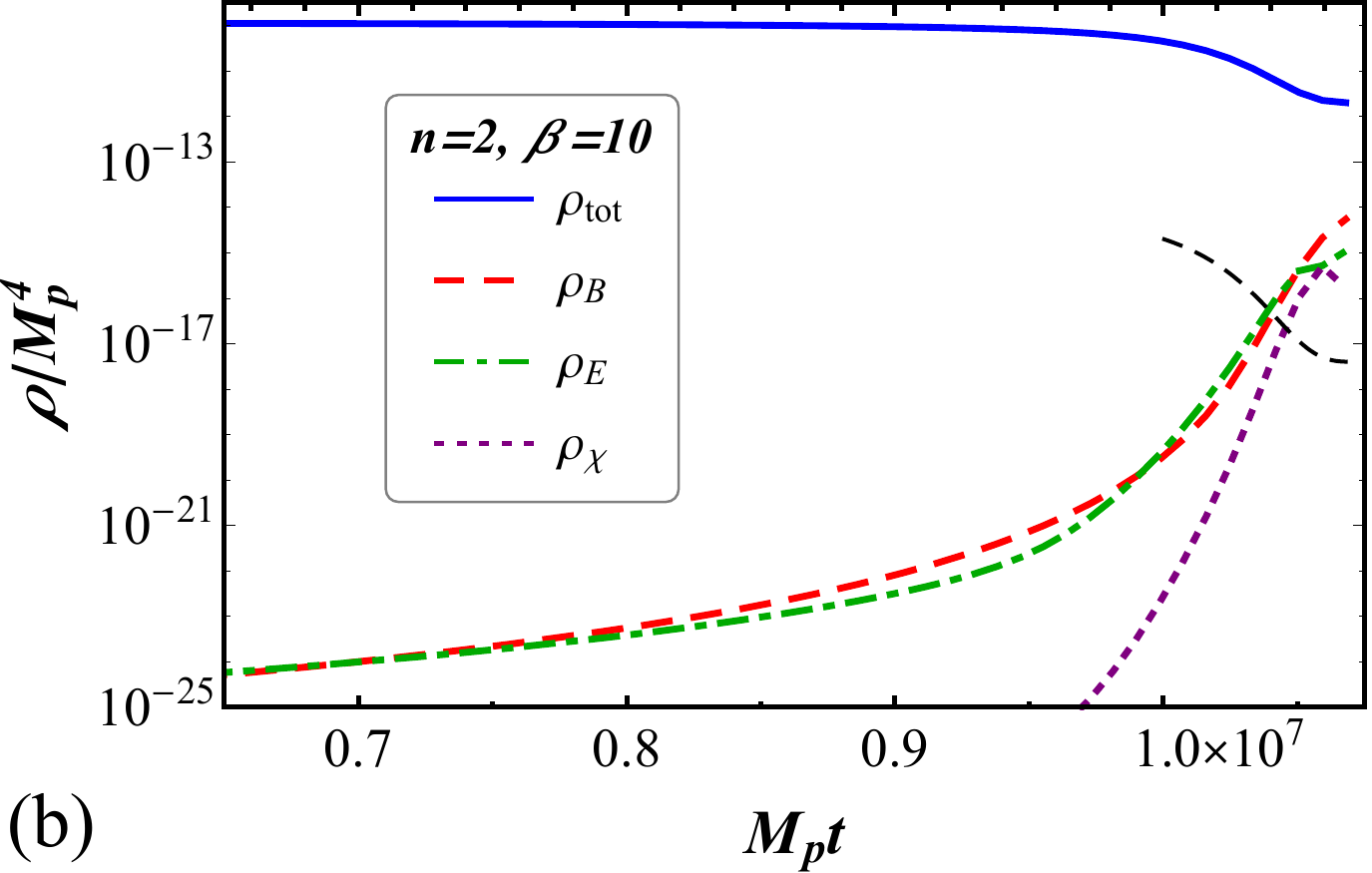}
	\caption{Panel (a): the time dependence of the EM energy density for the coupling constant $\beta=10$ calculated using three different approximation schemes: $n=0$ (purple dotted line), $n=1$ (green dashed-dotted line), and $n=2$ (red dashed line). For comparison, the blue solid line shows the result obtained by numerical integration of Eq.~(\ref{eq-mode-conformal}) for all relevant modes without taking into account the backreaction and Schwinger effect. The inset shows in more detail the region close to the end of inflation. Panel (b): the time dependence of the total energy density (blue solid line), the magnetic energy density (red dashed line), the electric energy density (green dashed-dotted line), and the energy density of charged particles generated due to the Schwinger effect (purple dotted line) calculated in the approximation scheme with $n=2$ for $\beta=10$. The thin black dashed line shows expression (\ref{condition-Schwinger}) which defines the electric energy density when the Schwinger effect becomes important. \label{fig-2-b10}}
\end{figure}

Now, let us study how the results depend on the order of the approximation scheme $n$ and how different components of the energy 
density evolve with time. The corresponding results are shown in Fig.~\ref{fig-2-b10} for $\beta=10$ for which the backreaction is irrelevant. Panel (a) compares different approximation schemes. The blue solid line corresponds to the exact solution of the mode equation (\ref{eq-mode-conformal}) with the Bunch-Davies initial condition (\ref{Bunch-Davies-vacuum}) in the assumption that the backreaction of the generated fields and the Schwinger effect do not affect the background evolution. The purple dotted line, green dashed-dotted line, and red dashed line show the EM energy density calculated using the approximation schemes with $n=0$ (equipartition between electric and magnetic energy densities), $n=1$, and $n=2$, respectively. First of all, it is very important that during the whole inflation stage all three used approximation schemes give results very close to the solution obtained by the numerical integration of Eq.~(\ref{eq-mode-conformal}) without taking into account the backreaction and Schwinger effect, see the main plot in Fig.~\ref{fig-2-b10}(a). The inset shows the time interval close to the end of inflation in more detail. Inspecting it, we conclude that higher approximation schemes are closer to the exact solution. We should note, however, that the convergence is asymptotic and higher order approximation schemes with $n\geq 3$ deviate from the exact solution at late times. 
Perhaps this can be explained by a large number of equations ($3n+5$) which we have to deal with when applying the scheme with large $n$ as the computational errors increase rapidly with $n$.

Figure~\ref{fig-2-b10}(b) shows the different components of the energy density calculated in the approximation scheme with $n=2$. 
The red dashed line gives the magnetic energy density, the green dashed-dotted line shows the electric energy density, the purple 
dotted line corresponds to the energy density of charged particles created due to the Schwinger effect. Finally, the blue solid line gives the 
total energy density. We see that the EM energy density is almost equally distributed between its electric and magnetic parts. This 
was observed previously in numerical lattice simulations performed in Refs.~\cite{Fujita:2015,Figueroa:2018}. 
At the beginning of inflation, the evolution of the EM field is determined by the competition of two processes: the Hubble damping due to the Universe expansion [the second term on the left-hand side of Eqs.~(\ref{eq-E-n})--(\ref{eq-B-n})] and enhancement due to new modes crossing the horizon [the boundary terms on the right-hand side of Eqs.~(\ref{eq-E-n})--(\ref{eq-B-n})]. Since for $\xi \sim 1$ the magnetic boundary term is larger than the electric one, see Eqs.~(\ref{bound-E}) and (\ref{bound-B}), and the Hubble damping has the same form, the ``equilibrium'' value of the magnetic energy density is larger than the electric one. However, close to the end of inflation, the electric energy density starts growing faster and crosses the magnetic energy density curve, see Fig.~\ref{fig-2-b10}(b). This growth can be explained by a faster rolling of the inflaton and by the presence of the term $-2I'(\phi)\dot{\phi}\mathscr{G}^{(0)}$ in Eq.~(\ref{eq-E-n}) for $n=0$ which has the negative sign and, thus, increases the time derivative. Note that such a term is absent in Eq.~(\ref{eq-B-n}) for the magnetic counterpart. The second crossing of these curves is explained by a fast decrease of the electric energy density due to the Schwinger effect coming into play at the end of inflation.
The energy density of charged 
particles is negligibly small compared to that of the EM field during almost the whole inflation stage. Only close to the end of inflation, the 
electric field becomes strong enough to satisfy condition (\ref{condition-Schwinger}) whose lower bound is shown in Fig.~\ref{fig-2-b10}(b) by 
the thin black dashed line. After that, the Schwinger effect produces the charged particles more intensively and their energy density 
quickly becomes comparable to that of the EM field.

\begin{figure}[h!]
	\centering
	\includegraphics[height=5.1cm]{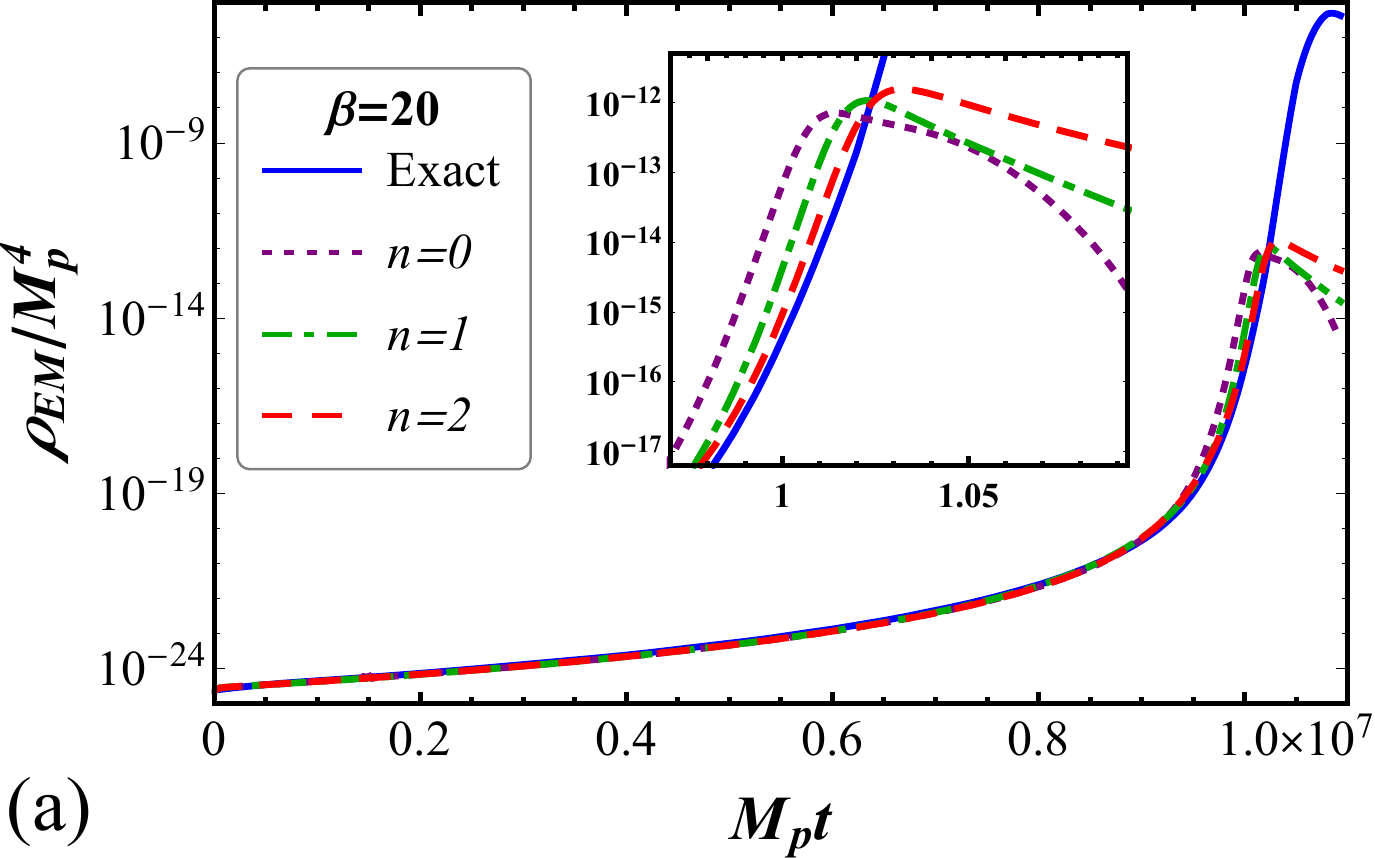} \hfill
	\includegraphics[height=5.1cm]{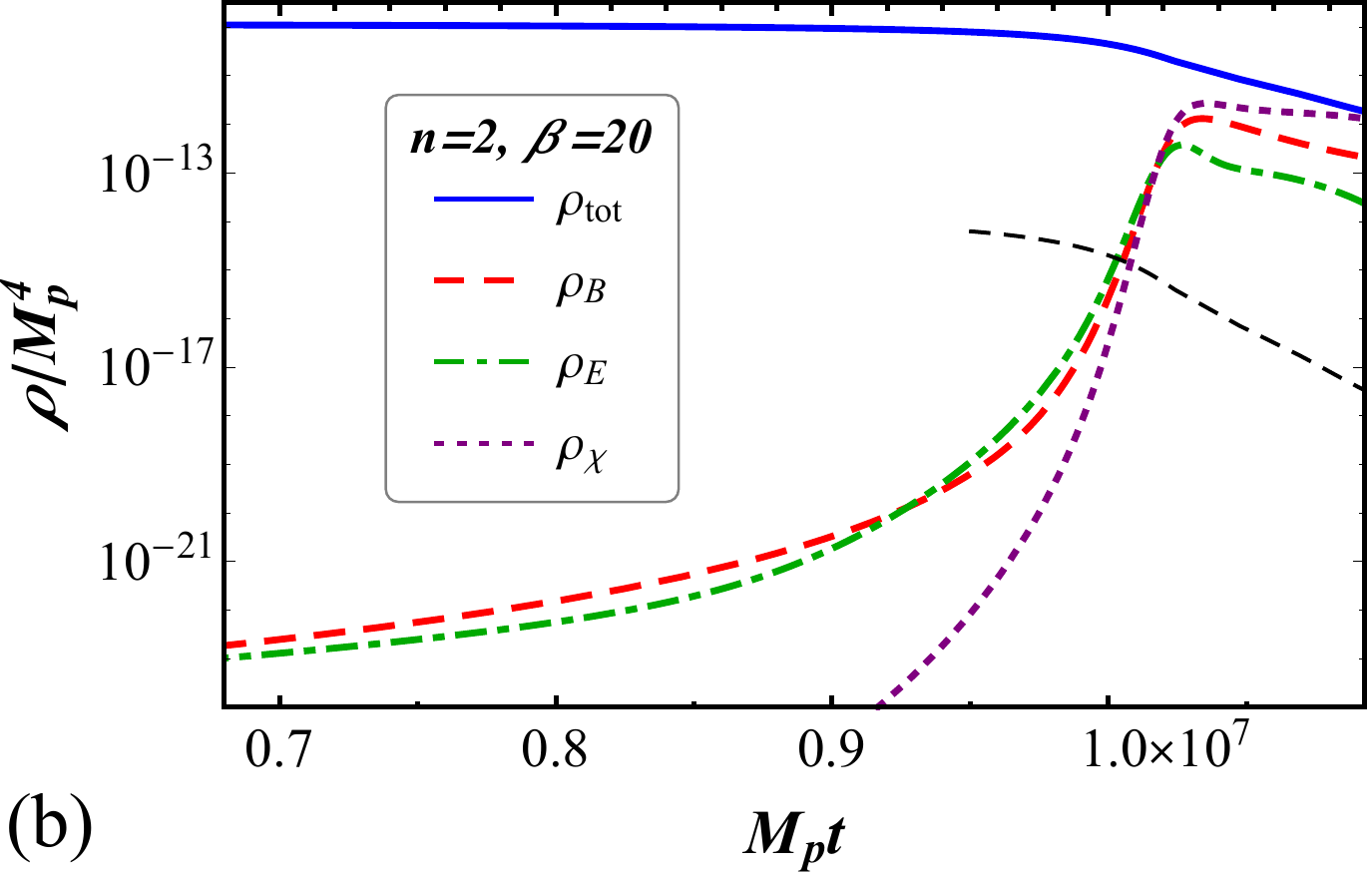}
	\caption{The same as shown in Fig.~\ref{fig-2-b10} for the coupling constant $\beta=20$. \label{fig-3-b20}}
\end{figure}

Further, let us consider larger values of the coupling constant. The plots similar to Fig.~\ref{fig-2-b10} are shown in Fig.~\ref{fig-3-b20} for $\beta=20$. According to panel (a), the backreaction becomes important close to the end of inflation. Consequently, the EM energy density suddenly stops growing and deviates from the result obtained neglecting the backreaction (blue solid line). As in the case of small $\beta$, higher order approximation schemes are in better correspondence with the exact solution, however, only for small $n$. 
Finally, Fig.~\ref{fig-3-b20}(b) shows that the Schwinger effect comes into play a bit earlier
than for $\beta=10$ and produces much more charged particles. Close to the end of inflation, their energy density becomes larger than that of the EM field and, due to the decrease of the inflaton energy density, it finally becomes the dominant part of the energy density of the Universe. Thus, the Schwinger effect helps to fill the Universe with charged particles even before the stage of inflaton fast oscillations. In the literature, this is known as the ``Schwinger preheating'' scenario \cite{Tangarife:2017,Sobol:2018}.

\begin{figure}[h!]
	\centering
	\includegraphics[height=5.1cm]{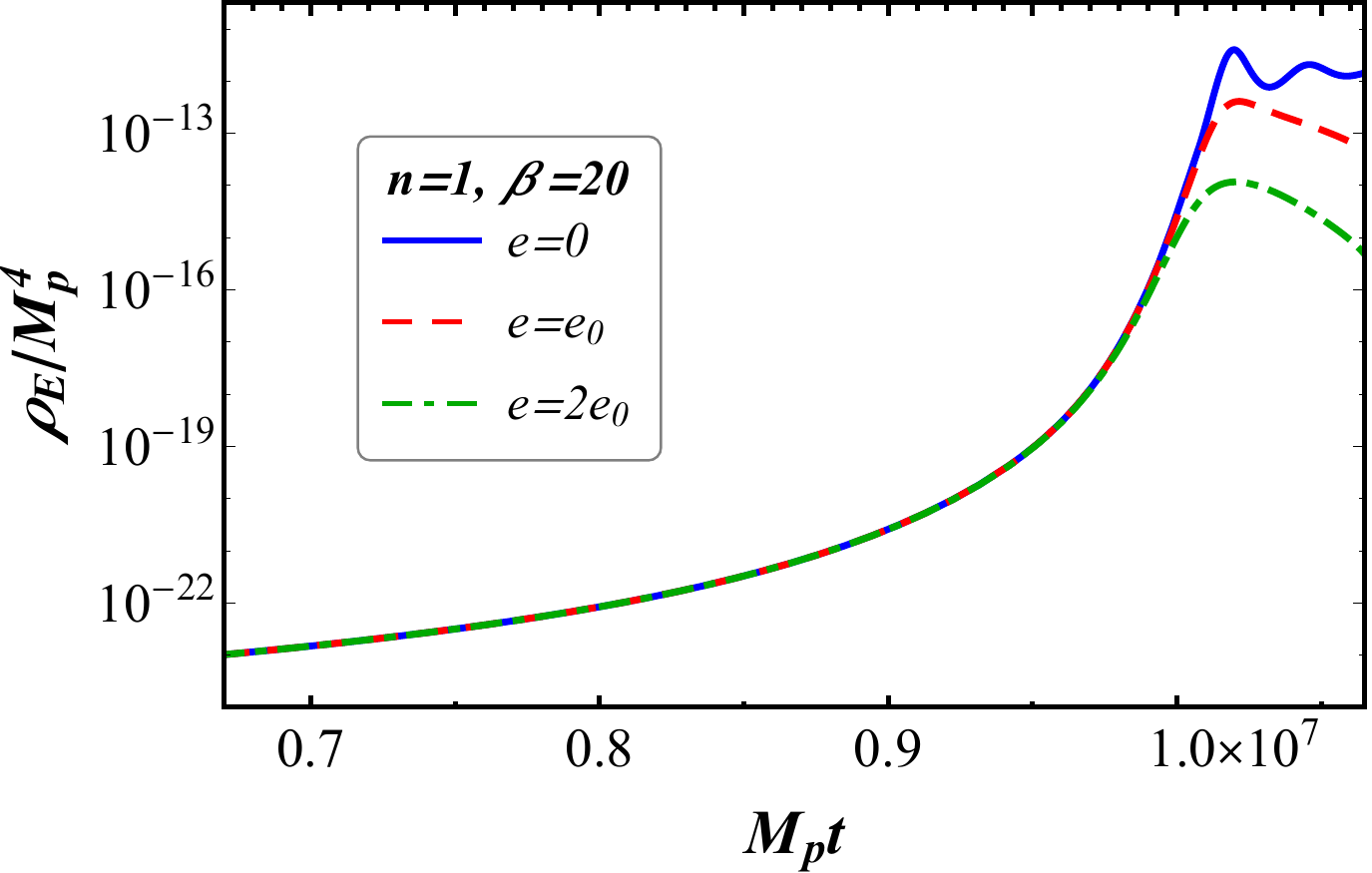}
	\caption{The time dependence of the electric energy density calculated in the approximation scheme $n=1$ for the coupling constant $\beta=20$ and three values of the charge of spin-$1/2$ particles: $e=0$ when the Schwinger effect is turned off (blue solid line), $e=e_{0}$ (red dashed line), and $e=2e_{0}$ (green dashed-dotted line). Here $e_{0}=\sqrt{4\pi\alpha_{\rm EM}}$ is the absolute value of the electron charge. \label{fig-4-Schwinger}}
\end{figure}

We also studied how the Schwinger effect influences the electric field. Figure~\ref{fig-4-Schwinger} shows the time dependence of the electric energy density for three values of the electric charge of created particles. The blue solid line corresponds to the case without the Schwinger effect ($e=0$). Here, after entering the backreaction regime, the electric energy density remains almost constant with small oscillations. For nonzero electric charge, the Schwinger effect turns on near the end of inflation and suppresses the electric energy density compared to the case $e=0$. This suppression is stronger and occurs earlier for larger values of the electric charge. Indeed, the lower bound for the electric energy density (\ref{condition-Schwinger}), when the Schwinger effect becomes important, depends on the electric charge as $e^{-6}$. Therefore, increasing the charge twice should suppress the electric field by $2^6=64$, which is in nice correspondence with the results in Fig.~\ref{fig-4-Schwinger}.

\begin{figure}[h!]
	\centering
	\includegraphics[height=5.1cm]{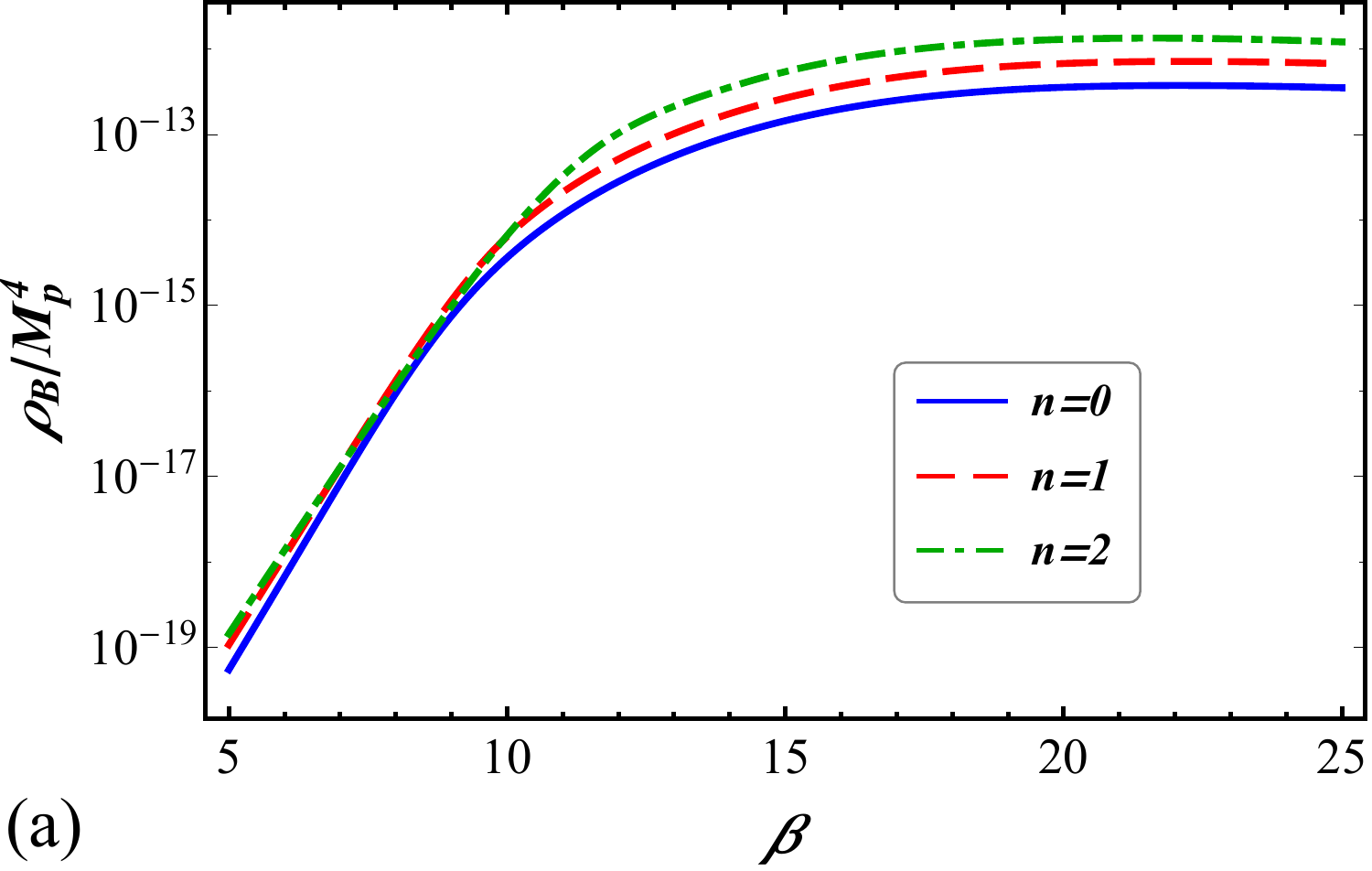} \hfill
	\includegraphics[height=5.1cm]{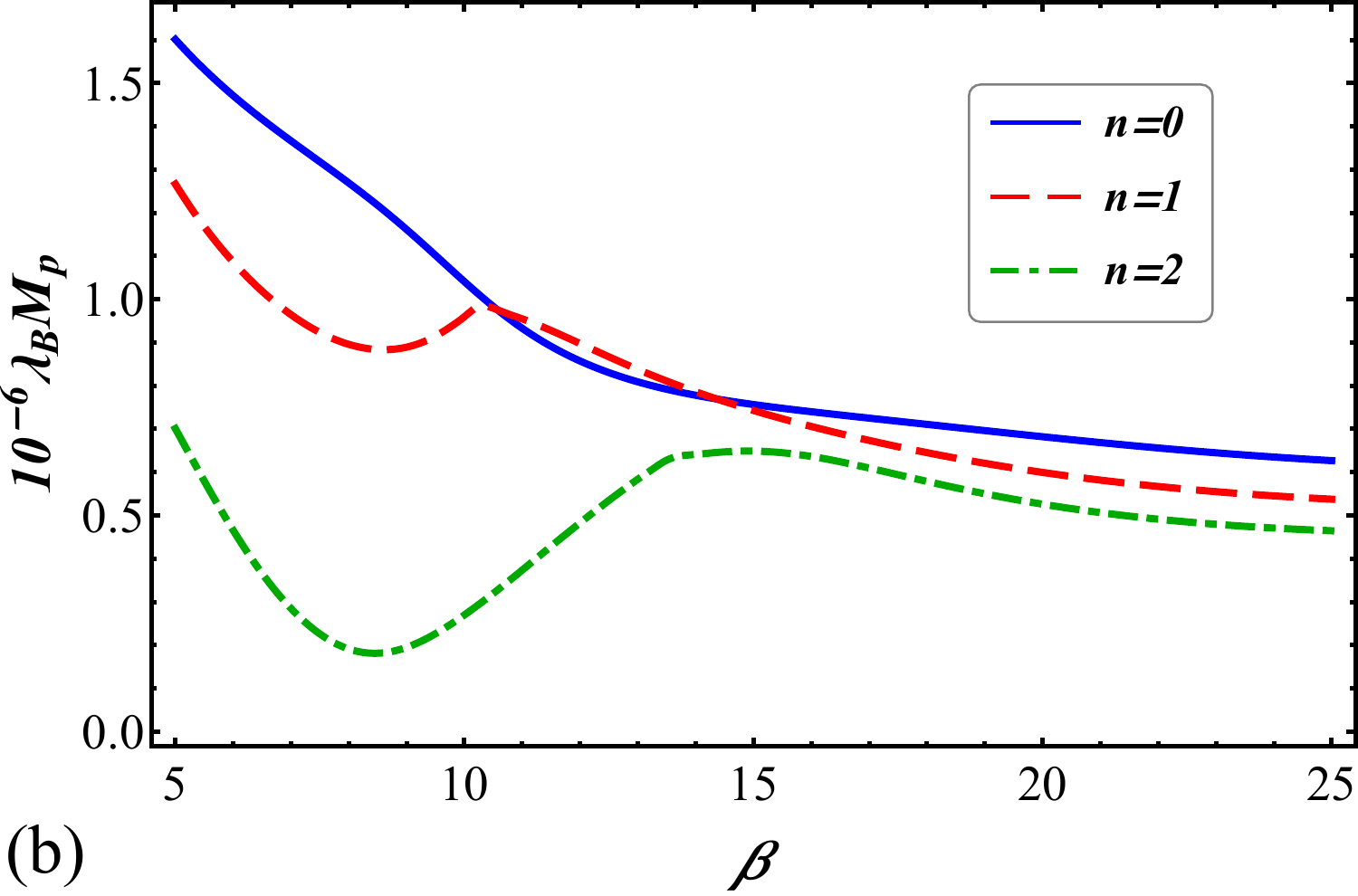}
	\caption{The dependence of the energy density (a) and the correlation length (b) of the generated magnetic field on coupling constant $\beta$ at the end of inflation calculated by using three approximation schemes: $n=0$ (blue solid line), $n=1$ (red dashed line), and $n=2$ (green dashed-dotted line).  \label{fig-5-mag-field}}
\end{figure}

Finally, we calculated the magnetic energy density and the magnetic correlation length at the end of inflation. These quantities, determined in three approximation schemes with $n=0$, 1, and 2, are given in Fig.~\ref{fig-5-mag-field} as functions of the coupling constant $\beta$. Panel (a) shows that the magnetic energy density at first grows with $\beta$ and then starting from $\beta\sim 10$ approaches a constant value $\rho_{B}\sim 10^{-12} M_{p}^{4}$, which slightly depends on the applied approximation scheme. The latter behavior can be explained by the fact that for $\beta >10$ the backreaction regime occurs prior to the end of inflation and the growth of the EM field terminates. Thus, for large $\beta$, the backreaction imposes a universal upper bound for the magnetic energy density generated during inflation which does not depend on $\beta$. The correlation length $\lambda_{B}$ always has the same order of magnitude, namely, $\lambda_{B}\sim 10^{6}M_{p}^{-1}$, which is of the same order as the horizon at the end of inflation $\lambda_{H}=1/H_{e}$. For large values of $\beta$, all approximation schemes give almost the same results recovering the universal behavior in the backreaction regime. However, for small $\beta$, different approximation schemes produce the results differing by a factor $2$ to $5$. The correlation scale (\ref{corr-length}) is proportional to the ratio of $\mathscr{B}^{(0)}$ to $\mathscr{B}^{(-1)}$ and changes in a small range. Thus, even relatively small differences in $\mathscr{B}^{(n)}$ in different approximation schemes [see, e.g., Figs.~\ref{fig-2-b10}(a) and \ref{fig-3-b20}(a)] lead to visible deviations in $\lambda_{B}$.

\section{Postinflationary evolution of the magnetic field}
\label{sec-postinflationary}

Another important issue in the study of magnetogenesis is to determine how magnetic fields generated during inflation evolve through the sequence of cosmological epochs until the present. After the end of inflation, the preheating takes place when the inflaton field oscillates in the potential minimum and decays into different particles. The EM field could be amplified during the preheating stage due to the interaction with the rapidly oscillating inflaton, e.g., through the mechanism of parametric resonance \cite{Fujita:2015,Adshead:2015,Adshead:2016,Deskins:2013}. However, our method is inapplicable at this stage for several reasons: (i) the boundary terms were derived for the inflationary background when new modes cross the horizon, while during the preheating stage they reenter the horizon; (ii) the expressions for the Schwinger current	were derived for the de Sitter-like background where $H$ is almost constant; (iii) finally, our approximation scheme assuming $\mathscr{E}^{(n)}=\mathscr{B}^{(n)}$ would break down because the Schwinger effect suppresses the electric component compared to the magnetic one.
Moreover, the backreaction regime, which is favorable for obtaining large magnetic fields, slows down the inflaton evolution decreasing its oscillation amplitude and reducing the efficiency of the parametric resonance.
Another source of the magnetic field amplification could be the power exchange with the electric field due to the Faraday law \cite{Kobayashi:2019}. However, this mechanism requires electric fields much stronger than the magnetic ones, which is not the case in the problem under consideration. Therefore, we will assume in what follows that the magnetic field is not enhanced during the preheating stage.

Further, during reheating, the plasma of created particles thermalizes and the Universe enters its hot radiation dominated phase. We will assume that at this stage the turbulent regime does not occur and the magnetic field and its correlation length evolve adiabatically, i.e., the corresponding comoving quantities remain constant:
\begin{equation}
\label{comoving-quantities}
\tilde{B}(t_{\rm reh})=\tilde{B}(t_{e})=B_(t_{e})\left(\frac{a_{0}}{a_{e}}\right)^{-2}, \quad \tilde{\lambda}_{B}(t_{\rm reh})=\tilde{\lambda}_{B}(t_{e})=\lambda_{B}(t_{e})\frac{a_{0}}{a_{e}},
\end{equation}
where the rescaling factor $a_{0}/a_{e}$ is determined below.

There are many models describing reheating in the literature, see Refs.~\cite{Kofman:1997,Bassett:2006,Allahverdi:2010,Frolov:2010,Amin:2015} for a review. However, the lack of observational data does not allow one to discriminate between them. Nevertheless, this stage can be described phenomenologically by two parameters such as the temperature at the end of reheating $T_{\rm reh}$ and the effective equation of state $\bar{w}$ which is defined as follows \cite{Ringeval:2010,Martin:2014}:
\begin{equation}
\bar{w}\equiv\frac{1}{N_{\rm reh}-N_{e}}\int_{N_{e}}^{N_{\rm reh}} w(N)dN,
\end{equation}
where $w(N)=p/\rho$ is the current value of the equation of state parameter and $N_{e}$ and $N_{\rm reh}$ are the numbers of $e$-foldings at the end of inflation and reheating, respectively. 
Using the energy conservation law
\begin{equation}
\label{covatiant-en-conserv}
\dot{\rho}+3H(\rho+p)=0,
\end{equation}
we can relate the scale factors and energies at the end of inflation and reheating through $\bar{w}$ as
\begin{equation}
\frac{a_{\rm reh}}{a_{e}}=\left(\frac{\rho_{\rm reh}}{\rho_{e}}\right)^{-\frac{1}{3(1+\bar{w})}}.
\end{equation}
The energy density at the end of inflation depends on the inflationary model and in the absence of the backreaction of
EM fields can be expressed in terms of the inflaton potential
\begin{equation}
\label{rho-e}
\rho_{e}=\frac{3}{2}V_{e}.
\end{equation}
Indeed, using Eq.~(\ref{covatiant-en-conserv}) and the Hubble equation $H^{2}=\rho/(3M_{p}^{2})$, we find the following relation 
between pressure and the energy density:
\begin{equation}
p=\rho \left(\frac{2}{3}\epsilon_{1}-1\right),
\end{equation}
where $\epsilon_{1}$ is the first Hubble flow parameter, see Eq.~(\ref{Hubble-flow}). Taking into account that $\epsilon_{1}=1$ when 
inflation ends, we obtain $p_{e}=-\rho_{e}/3$ which implies Eq.~(\ref{rho-e}). Moreover, Eq.~(\ref{rho-e}) can be also used as an
order-of-magnitude estimate of the energy density at the end of inflation even in the case when the backreaction occurs, because, as we
see from the numerical results in, e.g., Fig.~\ref{fig-3-b20}(b), the energy density is distributed between the inflaton, the EM field, 
and charged particles in comparable amounts.

Further, the CMB observations \cite{Planck:2018-infl} provide constraints on the amplitude and spectral properties of the primordial 
perturbations which, in turn, can be expressed through the inflaton potential at the moment when the pivot scale $k_{\ast}$ exits the Hubble 
horizon (in what follows we mark all quantities at this moment by asterisk). In particular, the amplitude of 
the power spectrum of scalar perturbations equals
\begin{equation}
\label{amplitude}
A_{s}=\left.\left(\frac{H^{2}}{2\pi|\dot{\phi}|}\right)^{2}\right|_{\ast}=\frac{1}{24\pi^{2}\epsilon_{\ast}}\frac{V_{\ast}}{M_{p}^{4}}.
\end{equation}
The tensor-to-scalar power ratio depends on the first slow-roll parameter according to $r=16\epsilon_{\ast}$. Combining these two 
equations, we find the corresponding value of the inflaton potential
\begin{equation}
V_{\ast}=\frac{3}{2}\pi^{2}A_{s}r M_{p}^{4}.
\end{equation}
Then the energy density at the end of inflation is given by
\begin{equation}
\rho_{e}=\frac{9}{4}\pi^{2}A_{s} r \frac{V_{e}}{V_{\ast}} M_{p}^{4},
\end{equation}
where $A_{s}\simeq 2.1\times 10^{-9}$ is the spectral amplitude of scalar perturbations. The 
tensor-to-scalar power ratio is bounded from above by $r<0.064$ \cite{Planck:2018-infl} and its typical value in the plateau models is of 
order $0.01$. We would like to note that $V_{e}/V_{\ast}\sim \mathcal{O}(0.1)$ in the majority of inflationary models except the case of the 
large-field inflation $V(\phi)\propto \phi^{n}$, which is strongly disfavored according to the results of the \textit{Planck}
mission \cite{Planck:2018-infl}. Then the typical value of the energy density at the end of inflation is
\begin{equation}
\rho_{e}\sim 10^{-10}M_{p}^{4}\simeq 3.3\times 10^{63}\,{\rm GeV}. 
\end{equation}

The energy density at the end of reheating can be expressed through the corresponding temperature as usual
\begin{equation}
\rho_{\rm reh}=\frac{\pi^{2}}{30}g_{\rm reh}T_{\rm reh}^{4},
\end{equation}
where $g_{\rm reh}$ is the effective number of relativistic degrees of freedom at that time. Using the entropy conservation 
relation, we can relate the scale factors at the end of reheating and the present time as follows:
\begin{equation}
\frac{a_{0}}{a_{\rm reh}}=\left(\frac{g_{\rm reh}}{g_{0}}\right)^{1/3}\frac{T_{\rm reh}}{T_{0}},
\end{equation}
where $T_{0}=2.725\,{\rm K}=2.35\times 10^{-13}\,{\rm GeV}$ is the present CMB temperature and
$g_{0}=2+3\times 2\times\frac{7}{8}\times\frac{4}{11}\approx 3.9$ is the effective number of relativistic degrees of freedom at present. Collecting all together, we obtain the rescaling factor
\begin{equation}
\label{rescaling-factor}
\frac{a_{0}}{a_{e}}=\rho_{e}^{\frac{1}{3(1+\bar{w})}}T_{\rm reh}^{\frac{3\bar{w}-1}{3(1+\bar{w})}}
g_{\rm reh}^{\frac{\bar{w}}{3(1+\bar{w})}}T_{0}^{-1}g_{0}^{-1/3}\left(\frac{\pi^{2}}{30}\right)^{-\frac{1}{3(1+\bar{w})}},
\end{equation}
which determines the comoving quantities (\ref{comoving-quantities}) at the end of reheating.

Further, in the radiatively dominated epoch, the plasma is highly dense and conductive with extremely large kinetic and magnetic Reynolds 
numbers. In this turbulent regime the helical magnetic fields undergo the inverse cascade. The comoving helicity conservation and the 
equipartition of energy between the kinetic motion in plasma and the magnetic field imply \cite{Banerjee:2004,Durrer:2013,Subramanian:2016}
\begin{equation}
	\tilde{B}\propto \tilde{\lambda}_{B}^{-1/2}\propto \eta^{-1/3},
\end{equation}
where $\eta$ is the conformal time. The spectrum of the magnetic field shifts to larger wavelength and lower magnitude and is self-similar all the time.

The inverse cascade takes place only in the turbulent regime which occurs when the Reynolds number is large. However, during the radiation domination epoch the turbulent phases alternate with the MHD viscous stages which include the periods of adiabatic evolution with $\tilde{\lambda}_{B}={\rm const}$ and the periods of rapid damping due to the free-streaming of the neutrinos or photons giving the major contribution to the viscosity. It was shown in Ref.~\cite{Banerjee:2004} that the evolution during the viscous stage gives approximately such values for the comoving magnetic field and its correlation length as if the viscous stage were absent at all. In other words, for numerical estimates of the comoving quantities, it is possible to assume that the magnetic field from the end of reheating until recombination evolves in the inverse cascade regime.

Taking into account that $\eta\propto a\propto T^{-1}$ during the radiation dominated stage and
$\eta\propto a^{1/2}\propto T^{-1/2}$ during the matter dominated epoch, we obtain the following comoving quantities at the moment of recombination:
\begin{equation}
\label{com-B-rec}
\tilde{B}(t_{\rm rec})=\tilde{B}(t_{e})\left(\frac{T_{\rm eq}}{T_{\rm reh}}\right)^{1/3}\left(\frac{T_{\rm rec}}{T_{\rm eq}}\right)^{1/6},
\end{equation}
\begin{equation}
\label{com-lambda-rec}
\tilde{\lambda}_{B}(t_{\rm rec})=\tilde{\lambda}_{B}(t_{e})\left(\frac{T_{\rm eq}}{T_{\rm reh}}\right)^{-2/3}\left(\frac{T_{\rm rec}}{T_{\rm eq}}\right)^{-1/3},
\end{equation}
where $T_{\rm eq}=6.54\times 10^{4} \Omega_{m0}h^2\,{\rm K}\simeq 9.16\times 10^{3}\,{\rm K}$ is the temperature at the matter-radiation equality and $T_{\rm rec}\simeq 4\times 10^{3}\,{\rm K}$ is the temperature at recombination.

If the correlation scale of the magnetic field becomes larger than the cosmic diffusion length $L_{\rm diff}\sim 1\,{\rm A.U.}$, then the magnetic field after recombination will undergo the free adiabatic evolution and Eqs.~(\ref{com-B-rec}) and (\ref{com-lambda-rec}) give the present day values of the magnetic field and its correlation length. On the contrary, if the correlation length is smaller than $L_{\rm diff}$, the modes with shorter wavelengths than $L_{\rm diff}$ decay after recombination. In this case, in order to find the present day value of the magnetic field, we should know the spectral index $n_{B}$ which parametrizes the magnetic power spectrum $d\rho_{B}/d\,\ln k \propto k^{n_{B}}$. Assuming that $n_{B}>1$ (this is correct for magnetogenesis in pseudoscalar inflation, see Refs.~\cite{Durrer:2011,Durrer:2013}), we obtain the following present day value of the magnetic field and its coherence length for $\tilde{\lambda}_{B}(t_{\rm rec})<L_{\rm diff}$:
\begin{equation}
\label{rescaling}
B(t_{0})=\tilde{B}(t_{\rm rec})\left[\frac{\tilde{\lambda}_{B}(t_{\rm rec})}{L_{\rm diff}}\right]^{\frac{n_{B}}{2}}, \quad \lambda_{B}(t_{0})=L_{\rm diff}.
\end{equation}
We will see below that for realistic values of the inflaton energy density and reheating temperature the correlation length at recombination appears to be larger than the cosmic diffusion scale and rescaling (\ref{rescaling}) is not needed.

Finally, using Eqs.~(\ref{comoving-quantities}), (\ref{rescaling-factor}), (\ref{com-B-rec}), and (\ref{com-lambda-rec}), we derive the general formulas for the present day magnetic field and its coherence length
\begin{equation}
\label{present-B}
B(t_{0})=\sqrt{2\rho_{B}(t_{e})}\left[\frac{30\rho_{e}}{\pi^{2}}g_{\rm reh}^{\bar{w}}T_{\rm reh}^{\frac{7\bar{w}-1}{2}}\right]^{-\frac{2}{3(1+\bar{w})}}g_{0}^{2/3}T_{0}^{2}(T_{\rm rec}T_{\rm eq})^{1/6},
\end{equation}
\begin{equation}
\label{present-lambda}
\lambda_{B}(t_{0})=\lambda_{B}(t_{e})\left[\frac{30\rho_{e}}{\pi^{2}}g_{\rm reh}^{\bar{w}}T_{\rm reh}^{5\bar{w}+1}\right]^{\frac{1}{3(1+\bar{w})}}g_{0}^{-1/3}T_{0}^{-1}(T_{\rm rec}T_{\rm eq})^{-1/3}.
\end{equation}

In order to estimate the numerical values of the present day quantities (\ref{present-B})--(\ref{present-lambda}), we consider two possible equations of state during reheating, namely, $\bar{w}=0$ if the preheating is described by decaying oscillations of the inflaton and $\bar{w}=1/3$ if the Schwinger effect is strong enough to transfer almost all energy density into ultrarelativistic charged particles or assuming that reheating is instantaneous. Choosing $10^{14}\,{\rm GeV}$ as a reference point for the reheating temperature, we obtain
\begin{equation}
B(t_{0})=
\left\{
\begin{array}{ll}
4.6\times 10^{-16}\,{\rm G}\left(\frac{\rho_{B}(t_{e})}{10^{-12}M_{p}^{4}}\right)^{1/2}\left(\frac{\rho_{e}}{10^{-10}M_{p}^{4}}\right)^{-2/3}\left(\frac{T_{\rm reh}}{10^{14}\,{\rm GeV}}\right)^{1/3}, & \text{for} \ \bar{w}=0,\\
4.1\times 10^{-15}\,{\rm G}\left(\frac{\rho_{B}(t_{e})}{10^{-12}M_{p}^{4}}\right)^{1/2}\left(\frac{\rho_{e}}{10^{-10}M_{p}^{4}}\right)^{-1/2}\left(\frac{T_{\rm reh}}{10^{14}\,{\rm GeV}}\right)^{-1/3}, & \text{for} \ \bar{w}=1/3,
\end{array}
\right.
\label{B-0}
\end{equation}
\begin{equation}
\lambda_{B}(t_{0})=
\left\{
\begin{array}{ll}
1.1\times 10^{-6}\,{\rm Mpc}\left(\frac{\lambda_{B}(t_{e})}{10^{6}M_{p}^{-1}}\right)\left(\frac{\rho_{e}}{10^{-10}M_{p}^{4}}\right)^{1/3}\left(\frac{T_{\rm reh}}{10^{14}\,{\rm GeV}}\right)^{1/3}, & \text{for} \ \bar{w}=0,\\
3.7\times 10^{-7}\,{\rm Mpc}\left(\frac{\lambda_{B}(t_{e})}{10^{6}M_{p}^{-1}}\right)\left(\frac{\rho_{e}}{10^{-10}M_{p}^{4}}\right)^{1/4}\left(\frac{T_{\rm reh}}{10^{14}\,{\rm GeV}}\right)^{2/3}, & \text{for} \ \bar{w}=1/3.
\end{array}
\right.
\label{lambda-0}
\end{equation}
Since the correlation scale is typically lower than $1\,{\rm Mpc}$, in order to compare our results with the data
of gamma-ray observations \cite{Neronov:2010,Tavecchio:2010,Taylor:2011,Dermer:2011,Caprini:2015}, we find the following expression for the 
effective field $B_{\rm eff}=\tilde{B}\sqrt{\tilde{\lambda}_{B}/(1\,{\rm Mpc})}$ by using Eqs.(\ref{B-0}) and (\ref{lambda-0}):
\begin{equation}
B_{\rm eff}(t_{0})=
\left\{
\begin{array}{ll}
4.8\times 10^{-19}\,{\rm G}\left(\frac{\rho_{B}(t_{e})}{10^{-12}M_{p}^{4}}\right)^{1/2}\left(\frac{\lambda_{B}(t_{e})}{10^{6}M_{p}^{-1}}\right)^{1/2}\left(\frac{\rho_{e}}{10^{-10}M_{p}^{4}}\right)^{-1/2}\left(\frac{T_{\rm reh}}{10^{14}\,{\rm GeV}}\right)^{1/2}, & \text{for} \ \bar{w}=0,\\
2.5\times 10^{-18}\,{\rm G}\left(\frac{\rho_{B}(t_{e})}{10^{-12}M_{p}^{4}}\right)^{1/2}\left(\frac{\lambda_{B}(t_{e})}{10^{6}M_{p}^{-1}}\right)^{1/2}\left(\frac{\rho_{e}}{10^{-10}M_{p}^{4}}\right)^{-3/8}, & \text{for} \ \bar{w}=1/3.
\end{array}
\right.
\end{equation}

We determined in the previous section the numerical values of the magnetic energy density and correlation length at the end of inflation in the $\alpha$-attractor model. By using them we estimate now the present day values of the magnetic field and its correlation scale. The energy density at the end of inflation is of order $\rho_{e}\sim 10^{-10}M_{p}^{4}$. In the most optimistic case, the magnetic energy density is of order $\rho_{B}(t_{e})\sim 10^{-12}M_{p}^{4}$ and the correlation scale is always $\lambda_{B}(t_{e})\sim 10^{6}M_{p}^{-1}$. Then, for the equation of state with $\bar{w}=0$, we obtain $B_{0}\simeq 5\times 10^{-16}\, {\rm G} \left(\frac{T_{\rm reh}}{10^{14}\,{\rm GeV}}\right)^{1/3}$ and $\lambda_{B,0}\simeq 10^{-6}\,{\rm Mpc} \left(\frac{T_{\rm reh}}{10^{14}\,{\rm GeV}}\right)^{1/3}$. The effective magnitude of the magnetic field is equal to $B_{\rm eff}\simeq 5\times 10^{-19}\, {\rm G} \left(\frac{T_{\rm reh}}{10^{14}\,{\rm GeV}}\right)^{1/2}$. For $\bar{w}=1/3$, the present day quantities are $B_{0}\simeq 4\times 10^{-15}\, {\rm G} \left(\frac{T_{\rm reh}}{10^{14}\,{\rm GeV}}\right)^{-1/3}$, $\lambda_{B,0}\simeq 4\times 10^{-7}\,{\rm Mpc} \left(\frac{T_{\rm reh}}{10^{14}\,{\rm GeV}}\right)^{1/3}$, and $B_{\rm eff}\simeq 2.5\times 10^{-18}\, {\rm G}$.

Higher values of the magnetic field are generated for larger reheating temperature $T_{\rm reh}$. However, the latter cannot exceed 
$T_{\rm max}\sim 10^{15}\,{\rm GeV}$ for which all energy density of the inflaton $\rho_{e}$ instantaneously transfers into radiation during 
reheating. In general, lower values of $T_{\rm reh}$ lead to lower $B_{0}$ except for $\bar{w}=1/3$. 
However, in this case, the correlation scale decreases with $T_{\rm reh}$ so that the effective strength of the magnetic field on the Mpc scale 
always remains of order $10^{-18}\,{\rm G}$. Therefore, even the most optimistic estimate can hardly explain 
the strength of the intergalactic magnetic fields from the distant blazar observations
\cite{Neronov:2010,Tavecchio:2010,Taylor:2011,Dermer:2011,Caprini:2015}.

\section{Conclusion}
\label{sec-concl}

\vspace{5mm}

In this work we investigated the influence of the backreaction of the EM field and the Schwinger effect on magnetogenesis in the pseudoscalar inflation model with the axial coupling of the EM field to the inflaton via the term $\beta (\phi/M_{p}) F_{\mu\nu}\tilde{F}^{\mu\nu}$. In order to perform a self-consistent study of this system, we extended the framework proposed in Ref.~\cite{Sobol:2018}, where magnetogenesis in the kinetic coupling model was studied. Its essence is a closed set of ordinary differential equations that describe the self-consistent evolution of classical observables in the form of quadratic functions of the electric and magnetic fields with an arbitrary number of the curls. The quantum origin of these observables is incorporated in the form of boundary terms in the corresponding evolution equations. These terms come from the contribution of EM modes which cross the horizon and undergo the quantum-to-classical transition transforming from the Bunch-Davies vacuum state to a state with large filling number.

The evolution equations for the expectation values of quadratic functions of the electric and magnetic fields contain terms with an extra power of the spatial curl. This results in an infinite chain of equations for the corresponding classical observables.  Therefore, some additional physical arguments should be used in order to terminate this system of equations to a finite set of equations. Since the coupling to the inflaton is symmetric with respect to the electric and magnetic fields, the physically reasonable approximation could be to set $\langle\mathbf{E}\cdot ({\rm rot})^{n}\mathbf{E}\rangle=\langle\mathbf{B}\cdot ({\rm rot})^{n}\mathbf{B}\rangle$ for a certain $n$. For $n=0$, this corresponds to an approximate equipartition of the energy density between the electric and magnetic components as found in the previous numerical studies \cite{Fujita:2015,Figueroa:2018} for small coupling constant. Going to higher order in $n$ seems to be a natural straightforward generalization.

The backreaction of the generated EM fields occurs when their energy density exceeds the value $\rho_{\rm EM}\sim (\sqrt{2\epsilon}/\beta) \rho_{\rm inf}$ slowing down the inflaton rolling along the slope of the potential. This makes the magnetogenesis less effective and the energy density of the EM field terminates the exponential growth of electromagnetic fields which remain almost constant until the preheating stage.

The generated strong electric fields lead to the charged particles production via the Schwinger process. Although there is also a particle production by the time-dependent metric of the expanding Universe, this contribution can be neglected in the strong field regime $\rho_{E}\gg H^{4}/\alpha_{\rm EM}$. Analytic expressions for the Schwinger current can be derived only in the simple case of a constant electric field in de Sitter background \cite{Kobayashi:2014,Froeb:2014,Bavarsad:2016,Stahl:2016a,Stahl:2016b,Hayashinaka:2016a,Hayashinaka:2016b,Sharma:2017,Bavarsad:2018,Hayashinaka:2018,Hayashinaka:thesis}. However, as was shown in Ref.~\cite{Geng:2018}, the form of the current for a time-dependent electric field in the strong field regime is the same as in the case of a constant field. We analyzed the importance of the Schwinger effect and found that it can affect magnetogenesis only if the inequalities $\rho_{E}\gtrsim (9\pi^{3}/8) H^{4}/\alpha_{\rm EM}^{3}\gg H^{4}/\alpha_{\rm EM}$ are satisfied, i.e., only in the strong field regime.

For numerical analysis, we used the $\alpha$-attractor model, which is a typical representative of the plateau models of inflation. The strength of the potential was chosen from the requirement of generation of the correct scalar power spectrum amplitude. We considered only the last 60 $e$-foldings of inflation although it can last much more. 
Taking this into account we imposed the initial conditions for the EM observables. Although the previous numerical studies \cite{Fujita:2015,Figueroa:2018} excluded the range of coupling constants $\beta>10$ in view of large inflaton perturbations and non-Gaussianities in the primordial spectra, we consider also large values of $\beta$ in order to test our method and to demonstrate more clearly the features of the backreaction regime.

We applied three approximation schemes with $n=0$, 1, 2, and compared their results with those found via the exact integration of the mode equation (\ref{eq-mode-conformal}), which does not take into account the backreaction and the Schwinger effect. For early times when the EM field is weak and does not cause the backreaction, all approximation schemes give the results which satisfactorily agree with the exact result. Close to the end of inflation, the backreaction takes place for $\beta>10$ and we observe the changes in the behavior of the EM energy density as predicted analytically. The Schwinger effect comes into play also only near the end of inflation. It suppresses the electric energy density and produces charged ultrarelativistic particles which can make the Universe radiation dominated even before the reheating stage. Indeed, for large values of the coupling constant $\beta\gtrsim 20$, the energy density of created particles becomes comparable with that of the inflaton at the end of inflation. This means that the Schwinger preheating should be also considered as an important complementary scenario to the standard one with the inflaton perturbative decay and parametric resonance.

Finally, we took into account the postinflationary evolution of the magnetic field and estimated its present day value. Because of the fact that the axial coupling to the inflaton enhances only one circular polarization of the EM modes, the generated magnetic fields have nontrivial helicity. Moreover, if the enhancement is significant, then the generated magnetic field is close to maximally helical because the second polarization may be neglected at all. The evolution of such a magnetic field is not described by the adiabatic decay due to the Universe expansion $B\propto a^{-2}$. In the turbulent plasma the magnetic field undergoes the inverse cascade when the magnetic energy is transferred from the short to large scales due to the helicity conservation. This helps to increase the correlation scale of the generated magnetic field and, as a result, to avoid the cosmic diffusion after the recombination stage. Taking the most optimistic estimate, the present day value of the magnetic field cannot exceed $10^{-15}$ G with the correlation length of 1 pc. This means that the effective magnetic field strength, which should be compared with the distant blazar observations, equals $B_{\rm eff}=B_{0}\sqrt{\lambda_{B,0}/(1\,{\rm Mpc})}\sim 10^{-18}\,{\rm G}$ and can hardly explain the observational data.

In this work we did not consider the inflaton and curvature perturbations generated by the EM field. Since the EM field affects the inflaton evolution, it can spoil the spectrum of the scalar and tensor perturbations probed by the CMB observations. This can also be used to constrain the parameter space of the model. However, the self-consistent inclusion of these perturbations into the system of equations would significantly complicate calculations and, therefore, it should be addressed elsewhere.

\begin{acknowledgments}
	The work of O.~O.~S. was supported by the ERC-AdG-2015 Grant No. 694896 and by the Swiss National Science Foundation Grant No. 200020B\_182864. 
	The work of S.~I.~V. was supported  by the Swiss National Science Foundation Grant No.  IZSEZ0\_186551.
	S.~I.~V. and O.~O.~S. are grateful to Professor~Mikhail Shaposhnikov for his kind hospitality at the Institute of Physics, \'{E}cole Polytechnique F\'{e}d\'{e}rale de Lausanne, Switzerland, where a part of this work was done.
	S.~I.~V. is grateful to Professor Marc Vanderhaeghen and Dr.~Vladimir Pascalutsa for their support and kind hospitality at the Institut f\"{u}r Kernphysik, Johannes Gutenberg-Universit\"{a}t Mainz, Germany, where a part of this work was done.
	The work of E.V.G. was supported partially by the Ukrainian State Foundation for Fundamental Research.

\end{acknowledgments}

\appendix

\section{Derivation of the initial conditions}
\label{App-math}

In this appendix we present the explicit expressions for the initial values of the EM quadratic functions (\ref{EE})--(\ref{BB}). They can be calculated in the approximation of an adiabatically slow change of the parameter $\xi$ defined in Eq.~(\ref{parameter-xi}). Substituting the approximate solution (\ref{solution-adiabatic-approximation}) into Eqs.~(\ref{EE-spectrum})--(\ref{BB-spectrum}), we obtained the initial conditions (\ref{EE-apr-2})--(\ref{BB-apr-2}) in the integral form as functions of $\xi_{0}$ which is the value of $\xi$ at the initial moment of time when the simulation starts.

The corresponding integrals can be calculated in terms of Meijer $G$ functions \cite{Luke:book}. We have
\begin{equation}
\label{EE-apr-4}
\mathscr{E}^{(n)}(0)
= \lambda^{n}\frac{H^{n+4}e^{\pi|\xi_{0}|}{\rm sh\,}(\pi|\xi_{0}|)}{2^{3n+11}\pi^{5/2}|\xi_{0}|^{n+3}}
G^{3,1}_{2,4}\left(16\xi_{0}^{2}
\left|
\begin{array}{c}
1; n+9/2\\
n+4,\,n+4,\,n+4;\,0
\end{array}
\right.
\right),
\end{equation}
\begin{equation}
\label{EB-apr-4}
\mathscr{G}^{(n)}(0)=\lambda^{n+1}\frac{H^{n+4}e^{\pi|\xi_{0}|}{\rm sh\,}(\pi|\xi_{0}|)}{2^{3n+13}\pi^{5/2}|\xi_{0}|^{n+4}}
G^{3,1}_{2,4}\left(16\xi_{0}^{2}
\left|
\begin{array}{c}
1; n+9/2\\
n+4,\,n+4,\,n+5;\,0
\end{array}
\right.
\right),
\end{equation}
\begin{equation}
\label{BB-apr-4}
\mathscr{B}^{(n)}(0)=\lambda^{n}\frac{H^{n+4}e^{\pi|\xi_{0}|}{\rm sh\,}(\pi|\xi_{0}|)}{2^{3n+15}\pi^{5/2}|\xi_{0}|^{n+5}}
G^{3,1}_{2,4}\left(16\xi_{0}^{2}
\left|
\begin{array}{c}
1; n+11/2\\
n+4,\,n+5,\,n+6;\,0
\end{array}
\right.
\right).
\end{equation}

For further convenience, we derive also the asymptotic expressions in the limiting cases of small and large $|\xi_{0}|$. 
For $|\xi_{0}|\ll 1$, we can expand the Macdonald functions to the leading order for $z\to 0$ and then the integrals can be 
easily calculated
\begin{eqnarray}
\label{EE-appr-small}
\mathscr{E}^{(n)}(0)
&\approx& \lambda^{n}\frac{H^{n+4} (2|\xi_{0}|)^{n+6}}{2\pi^{2}(n+4)^{3}}\left[2(n+4)^{2}\ln^{2}(2b|\xi_{0}|)-2(n+4)\ln(2b|\xi_{0}|)+1\right],\\
\mathscr{G}^{(n)}(0)
&\approx& \lambda^{n+1}\frac{H^{n+4} (2|\xi_{0}|)^{n+5}}{4\pi^{2}(n+4)^{2}}\left[1-2(n+4)\ln(2b|\xi_{0}|)\right],\\
\mathscr{B}^{(n)}(0)
&\approx& \lambda^{n}\frac{H^{n+4} (2|\xi_{0}|)^{n+4}}{4\pi^{2}(n+4)},
\end{eqnarray}
where $b=\exp(\gamma_{E})$ and $\gamma_{E}=0.577...$ is the Euler-Mascheroni constant. In the opposite case $|\xi_{0}|\gg 1$, the upper bound of 
integration in Eqs.~(\ref{EE-apr-2})--(\ref{BB-apr-2}) can be extended to infinity. Then by using Eqs.(\ref{int-1})--(\ref{int-3}), we 
obtain
\begin{eqnarray}
\label{EE-apr-3}
\mathscr{E}^{(n)}(0)&\approx& \lambda^{n}\frac{H^{n+4}e^{2\pi|\xi_{0}|}}{\pi^{3}|\xi_{0}|^{n+3}}\frac{[(n+3)!]^{4}}{2^{n+5}(2n+7)!},\\
\label{EB-apr-3}
\mathscr{G}^{(n)}(0)&\approx& \lambda^{n+1}\frac{H^{n+4}e^{2\pi|\xi_{0}|}}{\pi^{3}|\xi_{0}|^{n+4}}\frac{[(n+3)!]^{4}(n+4)}{2^{n+7}(2n+7)!},\\
\label{BB-apr-3}
\mathscr{B}^{(n)}(0)&\approx& \lambda^{n}\frac{H^{n+4}e^{2\pi|\xi_{0}|}}{\pi^{3}|\xi_{0}|^{n+5}}\frac{[(n+4)!]^{4}}{2^{n+7}(2n+9)!}
\frac{n+5}{n+4}.
\end{eqnarray}

We present here also the integrals used in the derivation of Eqs.~(\ref{EE-apr-3})--(\ref{EB-apr-3})
\begin{eqnarray}
\int_{0}^{\infty}z^{2n+1}K_{0}^{2}(z)dz&=&\frac{2^{n-1}(n!)^{3}}{(2n+1)!!},\label{int-1}\\
\int_{0}^{\infty}z^{2n+1}K_{1}^{2}(z)dz&=&\frac{2^{n-1}(n-1)!n!(n+1)!}{(2n+1)!!}, \label{int-2}\\
\int_{0}^{\infty}z^{2n}K_{0}(z)K_{1}(z)dz&=&\frac{2^{n-2}[(n-1)!]^{2}n!}{(2n-1)!!}.\label{int-3}
\end{eqnarray}

\end{document}